\documentclass[a4paper,fleqn,usenatbib]{mnras}

\usepackage{newtxtext,newtxmath}

\usepackage[T1]{fontenc}
\usepackage{ae,aecompl}

\usepackage{graphicx}	
\usepackage{amsmath}	
\usepackage[inline]{enumitem}
\usepackage{cleveref}
\usepackage{xcolor}
\usepackage{hyperref}

\newcommand{\matr}[1]{\mathsf{\mathbf{#1}}}

\title[Reduced redundant-baseline calibration]{Calibration schemes with $\mathcal{O}(N\log{N})$ scaling for large-N radio interferometers built on a regular grid}

\author[D. B. Gorthi et al.]{
Deepthi B. Gorthi,$^{1}$\thanks{E-mail: deepthigorthi@berkeley.edu (DBG)}
Aaron R. Parsons,$^{1}$
Joshua S. Dillon$^{1,\dagger}$
\\
$^{1}$Department of Astronomy, U. California, Berkeley, CA\\
$^{\dagger}$ NSF Astronomy and Astrophysics Postdoctoral Fellow\\
}

\date{Accepted XXX. Received YYY; in original form ZZZ}

\pubyear{2020}

\begin{document}
\label{firstpage}
\pagerange{\pageref{firstpage}--\pageref{lastpage}}
\maketitle

\begin{abstract}
    Future generations of radio interferometers targeting the 21\,cm signal at cosmological distances with $N\gg 1000$ antennas could face a significant computational challenge in building correlators with the traditional architecture, whose computational resource requirement scales as $\mathcal{O}(N^2)$ with array size. The fundamental output of such correlators is the cross-correlation products of all antenna pairs in the array. The FFT-correlator architecture reduces the computational resources scaling to $\mathcal{O}(N\log{N})$ by computing cross-correlation products through a spatial Fourier transform. However, the output of the FFT-correlator is meaningful only when the input antenna voltages are gain- and phase-calibrated. Traditionally, interferometric calibration has used the $\mathcal{O}(N^2)$ cross-correlations produced by a standard correlator. This paper proposes two real-time calibration schemes that could work in parallel with an FFT-correlator as a self-contained $\mathcal{O}(N\log{N})$ correlator system that can be scaled to large-N redundant arrays. We compare the performance and scalability of these two calibration schemes and find that they result in antenna gains whose variance decreases as $1/\log{N}$ with increase in the size of the array.
\end{abstract}




\section{Introduction} 
\label{sec:intro}

Traditional correlator architectures that have been used for most radio interferometers from the Very Large Array (VLA; \citealt{vla_1980}) to the Atacama Large Millimeter Array (ALMA; \citet{alma_correlator_2007}), require computational resources that scale as $\mathcal{O}(N^2)$ with the number of antennas. More recently, there has been a renewed interest in correlator architectures which require computational resources that scale less steeply with array size, for low-frequency radio astronomy applications that require a large collecting area. The Hydrogen Epoch of Reionization Array (HERA; \citealt{deboer_et_al2017}), the Canadian Hydrogen Intensity Mapping Experiment (CHIME; \citealt{bandura_et_al2014, newburgh_et_al2014}), the Murchison Widefield Array (MWA; \citealt{tingay_et_al2013}), LOw Frequency ARray (LOFAR; \citealt{vanhaarlem_et_al2013}) and MITEoR \citep{zheng_et_al2014} are all built with relatively cheap antennas that can scale to large-N arrays. At the low radio frequencies that these telescopes operate at, the signal chain can also be relatively inexpensive because cryogenic cooling of receivers is not essential. Receivers are sky-noise dominated at low radio frequencies \citep{Ellingson_2005}, decreasing the need to lower thermal-noise. If the correlator architecture can also scale up to large-N arrays, it will be more cost-efficient to build the collecting area required through numerous small antennas.

\begin{figure*}
    \centering \includegraphics[width=0.725\textwidth]{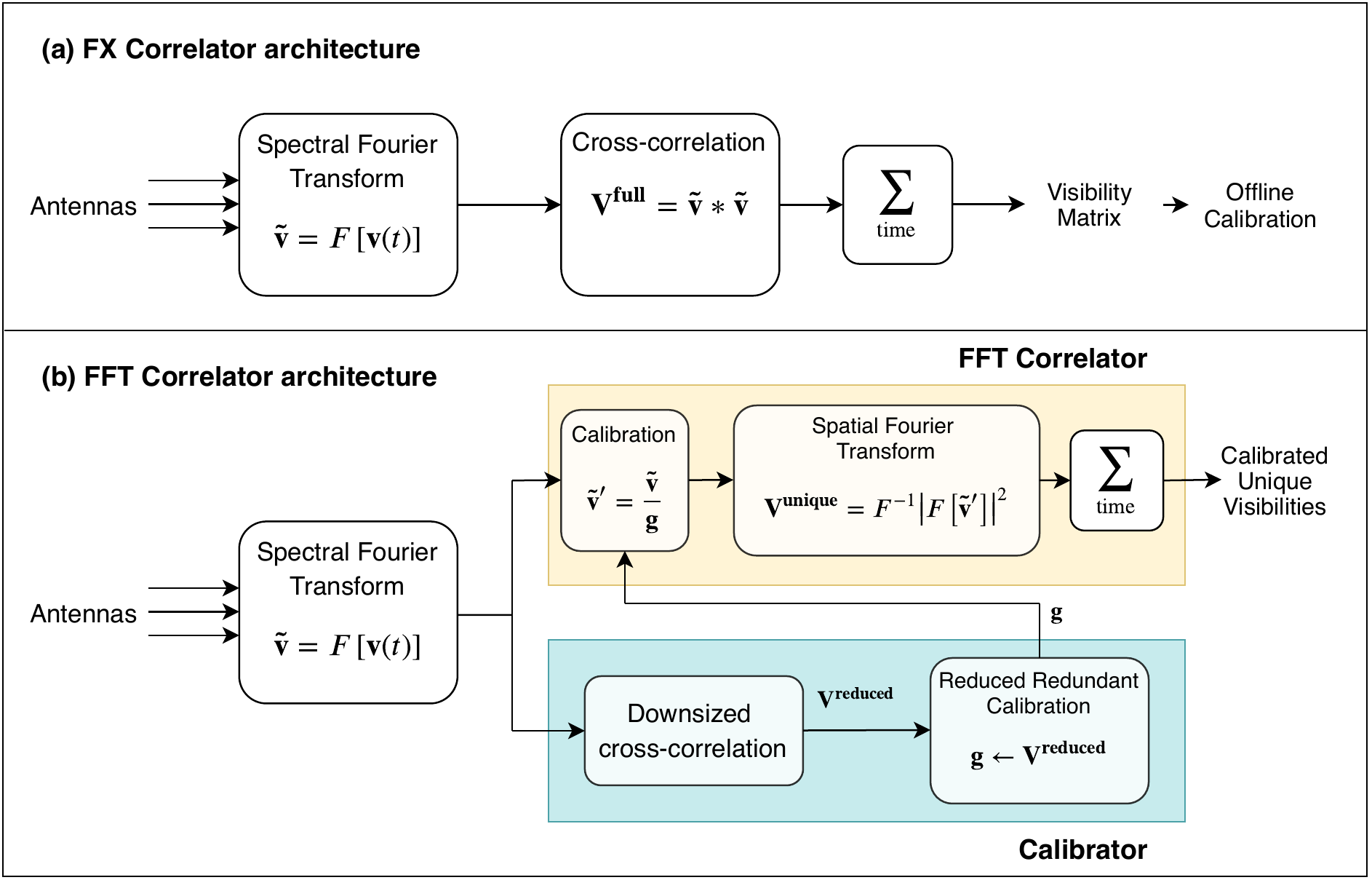}
    \caption{Panels (a) and (b) show the correlator architectures for a traditional FX-correlator and an FFT-correlator respectively. For either architecture, the first stage Fourier transforms the voltage measured by each antenna ${\bf v}(t)$ to obtain a spectrum ${\bf \tilde{v}}$, and the second stage computes visibilities of all antenna pairs. For an FX-correlator, the visibility matrix ${\bf V^{full}}$ is computed by cross-correlating the signal from every antenna with every other antenna in the array. For an FFT-correlator, the visibility matrix ${\bf V^{unique}}$ is computed by a spatial Fourier transform on the calibrated antenna voltages (yellow box). The calibrator (blue box) operates in parallel to the FFT-correlator and computes per-antenna gains for calibrating antenna voltages. The antenna gains are computed by performing one of the two reduced redundant-baseline calibration schemes described in this paper, on a smaller visibility matrix ${\bf V^{reduced}}$ computed for the purpose of calibration.
  \label{fig:systemlayout}}
\end{figure*}

In a traditional correlator architecture, the signal from every antenna is cross-correlated with the signal from every other antenna in the array. The product of individual cross-correlations are called \textit{visibilities} and the set of visibilities from different baselines in the array is called the \textit{visibility matrix}. Panel (a) of Figure~\ref{fig:systemlayout} shows the architecture of a traditional FX-correlator. The first stage performs a spectral Fourier transform, computing a spectrum of the time-varying voltage signal from antennas. The second stage computes the cross-correlation of all antenna pairs producing a time-integrated visibility matrix. The computational resources required to generate the visibility matrix and, to store and process the output data products scale as $\mathcal{O}(N^2)$ with the number of antennas in the array. For large-N arrays, this cost can dominate the entire cost of the array and has been one of the limiting factors for interferometers built in the previous decade.

For interferometers with antennas on a regular grid, \citet{daishido_et_al1991, tegmark_and_zaldarriaga2009, tegmark_and_zaldarriaga2010} have proposed FFT-correlators or FFT imagers as a potential solution to this steep scaling in cost- and computational-resources. Instead of cross-correlating antenna pairs, an FFT-correlator produces visibilities through a spatial Fourier transform. If the visibilities of redundant baselines, produced by an FX-correlator, can be averaged, these two methods are equivalent by the convolution theorem \citep[see][]{tegmark_and_zaldarriaga2009, tegmark_and_zaldarriaga2010}. By the nature of the Fast Fourier transform algorithm \citep{cooley_tukey65} FFT-correlators only scale as $\mathcal{O}(N\log{N})$, decreasing the number of computations performed in the correlator. 

An important difference between an FX-correlator and an FFT-correlator is that the latter does not preserve the full visibility matrix. The spatial Fourier transform averages redundant visibilities, which are expected to be the same, in principle, since they are visibilities measured by antenna pairs with the same displacement vector. However, in practice, redundant visibilities are different due to differences in the signal chain, structure of the dish, varying cable-lengths etc. If antenna voltages are not calibrated, the spatial Fourier transform could result in averaging dissimilar visibilities, making post-processing correction impossible as well. Hence, antenna gain- and phase-calibration, prior to the spatial Fourier transform, is essential to avoid signal-loss in the FFT-correlator.

An FFT-correlator that implements the design proposed by \citet{tegmark_and_zaldarriaga2009}, is the one built by \citet{foster_et_al2014} on the BEST-2 array at Medicina, Italy. They demonstrated that the visibilities produced by the FFT-correlator and the redundantly-averaged visibilities of an FX-correlator are similar when all the antennas are calibrated before the spatial Fourier transform. However, they used a traditional FX-correlator working in parallel to generate all the visibilities required for point-source calibration. This is not a scalable solution for calibrating large-N arrays since building an FX-correlator may not be viable. 

A more generic alternative to the FFT-correlator, discussed by \citet{thyagarajan_et_al2017}, is a direct-imaging-correlator called the E-field Parallel Imaging Correlator (EPIC) that has now been deployed on the LWA \citep{Kent_2019}. It works like a Modular Optimal Frequency Fourier (MOFF; \citealt{morales2011}) correlator, where antenna voltages are gridded before a spatial Fourier transform produces an electric-field image. Unlike the FFT-Correlator, EPIC can also be implemented on non-redundant arrays, including arrays where the antenna beams are non-identical. For highly redundant arrays with identical antenna beams, EPIC becomes equivalent to the FFT-correlator. \citet{beardsley2017} propose an iterative sky-based calibration algorithm, EPICal, for such a correlator that does not require generating real-time visibility products and scales as $\mathcal{O}(N)$. However, EPICal requires prior knowledge of antenna beams which can be difficult to model or measure in situ at low radio frequencies. Moreover, the lack of accurate diffuse-sky models that also account for polarisation at these frequencies could make it harder to decouple the sky-signal from beam models. In this paper, we choose to discuss only redundant array layouts where the redundancy can be exploited for calibration.

An ideal calibration scheme for FFT-correlators, must be capable of minimising the scatter in redundant visibilities because any residual scatter will become additional noise on the visibility returned by the correlator. Additionally, the calibration scheme must produce an output that can be applied to antenna voltages. \citet[\textsection~9]{Liu_and_Shaw_2019} summarise calibration methods that can be applied to redundant arrays. Redundant-baseline calibration \citep{wieringa1992, liu_et_al2010, noorishad_et_al2012, marthi_and_chengalur2014}, that has been used to calibrate the Donald C. Backer Precision Array for Probing the Epoch of Reionization (PAPER; \citealt{parsons_et_al2010, Ali_2015, Kolopanis_2019}), LOFAR \citep{noorishad_et_al2012}, MITEoR \citep{zheng_et_al2014, Zheng_et_al_2017} and HERA \citep{Dillon_et_al_2020}, results in complex antenna gains that can be applied to antenna voltages. The multiplicative antenna gains are computed by solving a system of equations that minimise the scatter in calibrated redundant visibilities.

A known caveat of redundant-baseline calibration is that it can only yield relative antenna gains \citep{liu_et_al2010, dillon_et_al2018} i.e, the equations can constrain the ratio of antenna gains but cannot determine their actual value. The system of equations has a null space with four degenerate parameters including the absolute amplitude and the phase of antenna gains. However, this is not a problem for calibrating voltages for the purpose of FFT-correlation since it requires only relative calibration of antennas so that visibilities of redundant baselines can be averaged coherently. Absolute calibration, to determine the degenerate parameters, can still be performed offline with the visibilities generated by the FFT-correlator. 

Applications of redundant-baseline calibration, so far, had the full visibility matrix available for constructing the system of equations for calibration. However, redundant-baseline calibration does not inherently require all $N(N-1)/2$ visibilities measured at high signal-to-noise ratio (SNR). This paper explores two redundant-baseline calibration schemes that can use $\mathcal{O}(N\log{N})$ computational resources for generating visibilities for the purpose of calibration and are henceforth referred to as \textit{reduced redundant-baseline calibration} schemes.

The FFT-correlator architecture assumed in this paper is similar to the one proposed by \citet{zheng_et_al2014}. This is shown in Panel (b) of Figure~\ref{fig:systemlayout}. The first stage, computing a spectrum of antenna voltages, is similar to the FX-correlator architecture. The yellow boxed region shows the FFT-correlation where a spatial Fourier transform on calibrated voltages results in the time-integrated unique visibilities of the array. Notice that the FFT-correlator does not produce the full visibility matrix; the redundant baselines are averaged by the spatial Fourier transform. 

The blue boxed region in Figure~\ref{fig:systemlayout} shows the \textit{calibrator}, which is the main focus of this paper. It performs two functions: (a) cross-correlate the baselines required for calibration in a manner similar to the second stage of an FX-correlator and (b) compute antenna gains by applying one of the two reduced redundant-calibration schemes on this set of visibilities. The computation- and resource-intensive stage of the calibrator is the first step of cross-correlating antenna pairs. The number of baselines that need to be cross-correlated in a given integration cycle determines the computational resources required by this stage of the calibrator. The reduced redundant-baseline calibration scheme employed by the calibrator dictates the set of visibilities that need to be cross-correlated and hence determines the size of the calibrator. Both the calibration schemes discussed in this paper can be adapted to a calibrator that scales as $\mathcal{O}(N\log{N})$, keeping the size of the calibrator comparable to the size of the FFT-correlator.

In the first reduced redundant-baseline calibration scheme, \textbf{low-cadence calibration}, the calibrator computes the full visibility matrix by cycling through baseline pairs. As an extreme example, if the computational resources allocated to the calibrator can only cross-correlate one antenna pair at a time, the low-cadence calibrator will generate the full visibility matrix by cycling through all the antenna pairs in the array. The full visibility matrix, constructed in this fashion, is then used for redundant-baseline calibration. While all the $\sim$$N^2/2$ visibilities are used for redundant-baseline calibration, they are computed by using only a small fraction of the resources of an $O(N^2)$ FX-correlator. By adjusting the integration time and the number of visibilities computed within each cycle, the computational resources required by a low-cadence calibrator can be limited to an $\mathcal{O}(N\log{N})$ scaling.

The second reduced redundant-baseline calibration scheme, \textbf{subset redundant calibration}\footnote{Shortened from subset redundant-baseline calibration}, is a generalisation of hierarchical redundant-baseline calibration described by \citet{zheng_et_al2014} (see Appendix~\ref{app:hierarcal}). In subset redundant calibration, the calibrator computes a partial visibility matrix by cross-correlating only a limited set of antenna pairs. Redundant-baseline calibration is applied to this partial visibility matrix to estimate the antenna gains. For example, in highly redundant arrays, since the shortest baselines involve all the antennas in the array, it is often possible to compute antenna gains by performing redundant-baseline calibration on just the shortest baselines. Since only a subset of the full visibility matrix is generated for the purpose of calibration, this technique is called subset redundant calibration. Depending on the baseline-types chosen for redundant-baseline calibration, the computational resources required by such a calibrator can also be limited to an $\mathcal{O}(N\log{N})$ scaling.

In the rest of this paper we attempt to show that the gains estimated using either low-cadence calibration or subset redundant calibration, are capable of minimising the scatter in the visibilities of redundant baselines. We lay out metrics for comparing the two reduced redundant-baseline calibration schemes and assessing their scalability to large-N arrays. Ultimately, we attempt to show that the calibrator design proposed here makes self-contained $\mathcal{O}(N\log{N})$ correlators conceivable for future generation large-N arrays.

The layout of the rest of the paper is as follows: Section~\ref{sec:redredcal} quantifies the parameters that are important for understanding the performance of either reduced redundant-baseline calibration method in our simulations (Section~\ref{sec:simulation}). Sections~\ref{sec:lowcadcal} and~\ref{sec:subredcal} examine low-cadence calibration and subset redundant calibration, respectively, and discuss the limits within which they result in convergent gain solutions. Section~\ref{sec:comparison} compares the performance of both methods for arrays of various sizes and discusses the limitations and advantages of employing either method for calibrating large-N arrays. Section~\ref{sec:conclusion} presents the conclusions of this paper.

\section{Metrics to Evaluate Reduced Redundant-Baseline Calibration}
\label{sec:redredcal}

Redundant-baseline calibration computes per-antenna complex gains by minimising the scatter in the visibilities of redundant baselines, which makes it a suitable calibration scheme for FFT-correlators. It relies on the fact that pairs of antennas with the same beam patterns, spaced at equal distances, measure the same visibility. If $V_{ij}$ is the visibility product of two antennas spaced a distance $d$ apart in the East-West direction, then the visibility measured by two different antennas, $V_{lm}$, is the same as $V_{ij}$ if they are also spaced a distance $d$ apart in the same direction. In practice, this is often not true because of variations in amplifier gain, timing differences originating in the correlator, cable delays etc., that need to be calibrated. By comparing visibilities that are theoretically identical, it is possible to infer the calibration parameters for the antennas involved.

In the case of highly redundant arrays such as HERA, PAPER, CHIME, and the MWA Phase-II hexes, there are many more visibility measurements than unique baselines. This allows one to build a system of equations, which can be solved to estimate all the antenna calibration parameters. For the array layout shown in Figure~\ref{fig:hexarray}, the system of equations can be constructed as\footnote{In general, all the measurements and variables in this system of equations have a time and frequency dependence. We have omitted writing this explicitly for notational convenience.}:

\begin{align}
    \label{eq:redcal}
    V_{01}^{\rm{meas}} &= g_0 g_1^* V_{\alpha}^{\rm{true}} + n_{01} \nonumber \\
    V_{12}^{\rm{meas}} &= g_1 g_2^* V_{\alpha}^{\rm{true}} + n_{12} \nonumber \\
&\vdots \quad (\text{baselines with }{\bf d_{\alpha}} = {\bf r_0} - {\bf r_1}) \nonumber \\
    V_{04}^{\rm{meas}} &= g_0 g_4^* V_{\beta}^{\rm{true}}  + n_{04} \nonumber \\
&\vdots \quad (\text{baselines with }{\bf d_{\beta}} = {\bf r_0} - {\bf r_4}) \nonumber \\
    V_{05}^{\rm{meas}} &= g_0 g_5^* V_{\gamma}^{\rm{true}} + n_{05} \nonumber \\
&\vdots \quad (\text{baselines with }{\bf d_{\gamma}} = {\bf r_0} - {\bf r_5}) \nonumber \\
    V_{02}^{\rm{meas}} &= g_0 g_2^* V_{\delta}^{\rm{true}} + n_{02} \nonumber \\
&\vdots \nonumber \\ 
\text{baseli}&\text{nes separated by one or more antennas}
\end{align}
\noindent
where $V_{\alpha}^{\rm{true}}$ is the unknown, model true visibility of all the baselines with a displacement vector ${\bf d_{\alpha}}$, $V_{\beta}^{\rm{true}}$ is the unknown model true visibility of baselines with displacement vector ${\bf d_{\beta}}$ and so on. Baselines with the same displacement vector are said to be of the same baseline-type. $V_{ij}^{\rm{meas}}$ is the visibility measured by the pair of antennas $(i,j)$ in the field, and $n_{ij}$ is the noise in that measurement. The per-antenna, complex gains denoted by $g_i$ represent the calibration parameters of the antennas involved in measuring that visibility. 

The redundant-baseline calibration process estimates the gains and models true visibilities that best describe the measured visibilities. When the full visibility matrix ${\bf V^{full}}$ is used for the set of measured visibilities $V_{ij}^{\rm{meas}}$, the $V_{\alpha}^{\rm{true}}$ returned by the redundant-baseline calibration process represents the minimum-scatter average visibility for that unique baseline-type.

The system of equations in Equation~\ref{eq:redcal} can also be built using the visibility matrix computed by the calibrator ${\bf V^{reduced}}$. In the case of low-cadence calibration, this set of visibilities may have a lower SNR than the full visibility matrix due to smaller integration times in the calibrator. In case of subset redundant calibration, this set of visibilities is smaller than the full visibility matrix (but sufficient to determine and over-constrain all the variables in the system of equations) due to fewer cross-correlations computed by the calibrator. The $V_{\alpha}$s estimated by either reduced redundant-baseline calibration schemes are discarded and only the antenna gains are used to calibrate voltages for the spatial Fourier transform. The redundant-baseline averaged visibilities for all unique baseline-types are computed by the FFT-correlator. 

For the purpose of this paper, it is useful to consider an intermediate hypothetical step in the spatial Fourier transform where the redundant baselines have not yet been averaged. At this stage, the visibilities in the FFT-correlator would be equivalent to the calibrated visibilities of an FX-correlator. The baseline averaging stage within the FFT-correlator, that generates ${\bf V^{unique}}$, can be written in terms of the full visibility matrix as:

\begin{equation}\label{eq:modelvis}
    V_{\alpha}^{\rm{unique}} = \frac{1}{N_{\alpha}}\sum_{(i,j) \in \alpha} \frac{V_{ij}^{\rm{full}}}{g_ig_j^*}
\end{equation}
\noindent
where $N_{\alpha}$ is the number of redundant baselines that contribute to that baseline-type. Since it is easier to quantify the effect of calibration on visibilities than on voltages, we use this equation to represent the process of calibration in the FFT-correlator.

Another difference between traditional redundant-baseline calibration and 
reduced redundant-baseline calibration, that is evident from Equation~\ref{eq:modelvis}, is that the latter estimates antenna gains from a different set of visibilities (${\bf V^{reduced}}$) than what they are finally applied to (${\bf V^{full}}$). In this section we discuss two metrics that will help in evaluating the effect of this: (a) the uncertainty in estimated antenna gains and (b) scatter in visibilities calibrated by a reduced redundant-baseline calibration process.

The uncertainty in antenna gains has to decrease or remain constant with increase in array size, for the reduced redundant-baseline calibration scheme, and consequently the calibrator design, to be scalable to large arrays. Moreover, the uncertainty in the estimated gains can be expressed in terms of the SNR of the measured visibilities and the number of baselines used in the calibration process, providing us with a convenient metric to directly compare low-cadence calibration and subset redundant calibration. The overall uncertainty in the redundant-baseline averaged calibrated visibilities comes from both the noise in the measured visibilities and the uncertainty in the estimated gains. A quantitative measure of the antenna gain uncertainty will help us estimate the contribution of gain errors to the overall error in the calibrated visibilities.

The scatter in visibilities post-calibration is a direct probe of the effectiveness of the reduced redundant-baseline calibration process in estimating gains that can calibrate antenna voltages for the FFT-correlation. The spatial Fourier transform averages the visibilities of redundant baselines (Equation~\ref{eq:modelvis}) converting any residual scatter into noise in the estimated visibilities. Hence, quantifying the post-calibration scatter will help  us evaluate the performance of either reduced redundant-baseline calibration scheme with respect to traditional redundant-baseline calibration.

An important mathematical detail, before delving into the metrics that assess reduced redundant-baseline calibration schemes, is that the system of equations represented by Equation~\ref{eq:redcal} is not linear. \citet{wieringa1992} suggests a logarithmic approach to linearizing which can be written as: 

\begin{equation}\label{eq:logcal}
\ln{V_{ij}^{\rm{meas}}} = \ln{g_i} + \ln{g_j^*} + \ln{V^{\rm{true}}_{\alpha}} + n^{\prime}_{ij}
\end{equation}
\noindent
The noise parameter $n^{\prime}_{ij}$, evaluates to a non-Gaussian error that depends on the SNR of the measured visibilities. \citet{liu_et_al2010} discuss the noise-bias in antenna gains due to this non-Gaussianity, and propose another approach based on Taylor expanding the variables around a starting point. This paper employs a widely used third approach, called \texttt{omnical} \citep{zheng_et_al2014, Ali_2015, li_et_al2018, Dillon_et_al_2020}, that was originally developed for the MITEoR experiment. We use the logarithmic approach to make theoretical arguments about the nature of gain solutions since constant coefficients make the system of equations easier to analyse. However, simulations run to test these arguments and the plots shown in this paper have been generated by employing the \texttt{omnical} algorithm that is not noise-biased. In general, most of the results presented in this paper are not dependent on the solving technique used.

\subsection{Uncertainty in Antenna Gains}
\label{sec:redredcal:gainvar}

The uncertainty in the antenna gains, estimated by solving a system of equations, is given by the variance-covariance matrix (covariance matrix henceforth). The linearized system of equations shown in Equation~\ref{eq:logcal} can be written in matrix notation as $\matr{A}\matr{x} + \matr{n} = \matr{b}$ where $\matr{A}$ is a constant complex-valued matrix of dimensions $N_m \times N_v$ ($N_m$ is the number of measured visibilities and $N_v$ is the number of variables). $\matr{x}$, $\matr{b}$ and $\matr{n^{\prime}}$ are one-dimensional matrices of the variables (log-gains and log-unique-visibilities), measured quantities (log-visibilities) and noise in each measurement respectively. The covariance matrix $\matr{C}$ (of dimensions $N_v \times N_v$) for the estimated variables is given by:

\begin{equation}\label{eq:cov}
\matr{C} = \left<\matr{x}\matr{x}^{\dagger}\right> = \left(\matr{A}^{\dagger}\matr{N}^{-1}\matr{A}\right)^{-1}
\end{equation}
\noindent
The diagonal of the covariance matrix gives the variance in estimated variables, including antenna gains. The noise covariance matrix ($\matr{N} = \left<\matr{n} \matr{n}^{\dagger}\right>$), is a statistical estimate of the noise in the measurement matrix $\matr{b}$. The matrix $\matr{n}$ cannot be measured and can only be estimated from the thermal noise expected in the measurements.

The covariance matrix $\matr{C}$ reflects the covariance between all variables returned by the reduced redundant-baseline calibration process. Since the calibrator uses only the antenna gains and discards the model visibilities estimated by redundant-baseline calibration, the covariance matrix of interest is the \textit{marginalised} covariance ($\matr{C^{\prime}}$) of just the antenna gain solutions. Assuming all variables are normally distributed, the marginalised covariance matrix for a subset of variables is given by the rows and columns of the variables of interest. Hence, the marginalised covariance matrix of the gain solutions is given by the first $N$ rows and columns of the covariance matrix in Equation~\ref{eq:cov}, where $N$ is the number of antennas in the array.

\citet[\textsection~2.4]{liu_et_al2010} derive the noise covariance matrix for the logarithmic approach to linearizing (Equation~\ref{eq:logcal}) under the assumptions that the measured visibilities have a high SNR, and that the noise in the measured visibilities is uncorrelated between baselines, Gaussian in nature and similar across all baseline-types. This noise covariance matrix evaluates to:

\begin{equation}
\label{eq:noisecov}
    \matr{N} \approx (\rm{SNR})^{-2} \; \matr{I}
\end{equation}
\noindent
under the additional assumption that all the baselines in the array have the same average SNR. In general, this assumption does not hold when observing a real sky. However, in this paper, we only use SNR in the context of other array-averaged parameters, for which this assumption is justified. Since we have assumed that the noise is uncorrelated between baselines, the noise covariance matrix is just proportional to the identity matrix $\matr{I}$.

Substituting the noise covariance matrix into Equation~\ref{eq:cov}, and taking the first $N$ rows and columns of the covariance matrix, which we denote by $\left(\matr{A}^{\dagger}\matr{A}\right)^{-1}_{(N\times N)}$, we get the covariance of antenna gains. The diagonal of this matrix represents the variance or uncertainty in the estimated antenna gains.

\begin{equation}
    \label{eq:gaincov}
    \sigma_{g}^2 \approx (\rm{SNR})^{-2} \; \mathrm{diag}\left[\left(\matr{A}^{\dagger}\matr{A}\right)^{-1}_{(\mathit{N}\times \mathit{N})}\right]
\end{equation}
\noindent
The two reduced redundant-baseline calibration schemes effect the uncertainty in gains according to the above equation. In low-cadence calibration, the lower SNR of visibilities in the calibrator result in higher gain variance as compared to traditional redundant-baseline calibration. In subset redundant calibration, only a sub-matrix of $\matr{A}$ is used for calibration, again resulting in antenna gains with a higher variance.

The matrix $(\matr{A}^{\dagger}\matr{A})$ is nearly diagonal (small off-diagonal terms), with each entry equal to the number of equations in which the corresponding variable is involved. When redundant-baseline calibration is performed using the full visibility matrix, the first $N$ diagonal entries of this matrix are each equal to $N$ since every antenna is involved in $N$ equations. At constant SNR, this results in the following scaling for gain variance:

\begin{equation}\label{eq:gainvar_redcal}
    \sigma_{g}^2 \propto \frac{1}{N}
\end{equation}
\noindent
In Sections~\ref{sec:lowcadcal} and~\ref{sec:subredcal} we derive a scaling relation for the variance in gains estimated using a $\mathcal{O}(N\log{N})$ calibrator that implements low-cadence calibration and subset redundant calibration respectively, and compare it with the above relation.

\subsection{Scatter in Visibilities of Redundant Baselines}
\label{sec:redredcal:scatter}

The spatial Fourier transform in an FFT-correlator, which decreases the computational scaling from $\mathcal{O}(N^2)$ to $\mathcal{O}(N\log{N})$, averages the visibilities of redundant baselines. Hence, the post-calibration residual scatter in redundant visibilities is an important metric for assessing the gains estimated by a reduced redundant-baseline calibration process.

The scatter in redundant visibilities is quantified by the reduced $\chi^2$ of antenna gains and model visibilities ($\chi_r^2$) which is given by:

\begin{equation}\label{eq:redchi}
    \chi_r^2 = \frac{1}{\rm{DoF}}\sum_{(i,j) \in \alpha, \forall \alpha} \frac{\left|V_{ij}^{\text{meas}} - g_i g_j^*V_{\alpha}^{\text{true}}\right|^2}{\sigma_{ij}^2}
\end{equation}
\noindent
where $\sigma^2_{ij}$ is the variance of the noise in measured visibilities, $n_{ij}$ in Equation~\ref{eq:redcal}. DoF is the degrees of freedom in this system of equations given by:

\begin{equation}\label{eq:dof}
    \rm{DoF} = \mathit{N}_{obs} - \mathit{N} - \mathit{N}_{ubl} + 2
\end{equation}
\noindent
$N_{\rm{obs}}$ is the total number of cross-correlations computed (or number of visibility measurements), $N$ is the number of antennas in the array and $N_{\rm{ubl}}$ is the number of unique baseline-types in the system of equations. The additional offset by two accounts for the number of degenerate parameters in the system of equations representing a single-polarisation \citep{dillon_et_al2018, Dillon_et_al_2020}.

The reduced redundant-baseline calibration process within the calibrator is setup to minimise the $\chi_r^2$ between the visibility matrix computed by the calibrator ${\bf V^{reduced}}$ and the estimated gains and model true visibilities. In an array with identical antennas and perfect redundancy, we expect this $\chi_r^2$ to evaluate to one.

The estimated gains, however, apply to antenna voltages prior to the spatial Fourier transform which computes a different visibility matrix ${\bf V^{unique}}$ than that used in the reduced redundant-baseline calibration process. Hence, the $\chi_r^2$ evaluated using the gains estimated by the reduced redundant-baseline calibration process, model visibilities computed by the FFT-correlator (estimated using Equation~\ref{eq:modelvis}) against the $V_{ij}^{\rm{meas}}$ drawn from the full visibility matrix computed by an FX-correlator, is a useful a metric to assess the effectiveness of the estimated gains in calibrating the whole array. 

In Section~\ref{sec:comparison}, we use the $\chi_r^2$ computed in this way to compare the performance of reduced redundant-baseline calibration to traditional redundant-baseline calibration with the full visibility matrix.

\section{Simulation} 
\label{sec:simulation}

\begin{figure}
  \centering \includegraphics[width=\linewidth]{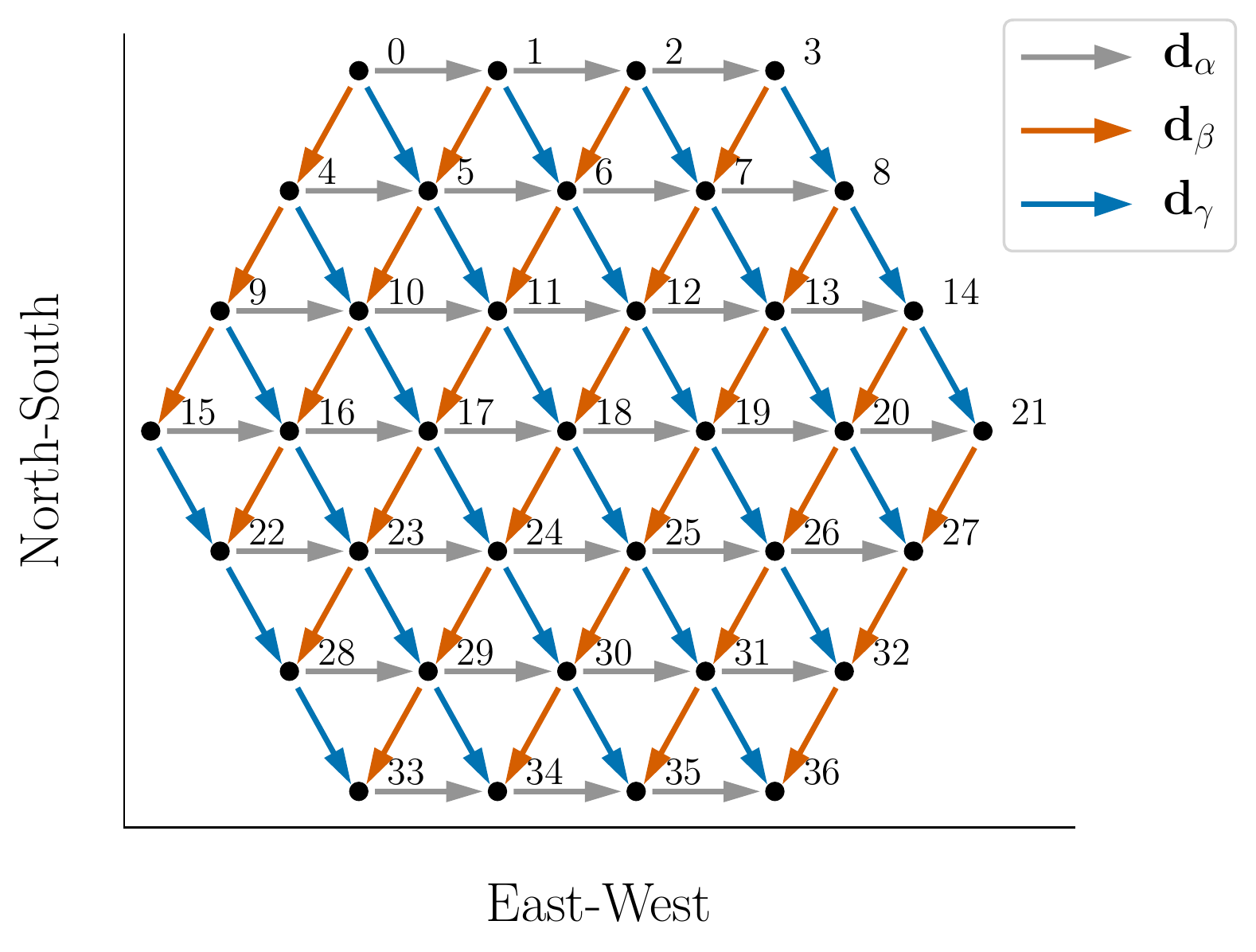}
  \caption{A hexagonal redundant array that is loosely based HERA; is used as a prototype for demonstrating the properties of reduced redundant-baseline calibration in this paper. Three different redundant baseline groups are marked by blue, orange and grey arrows.
    \label{fig:hexarray}}
\end{figure}

In the following sections, we discuss the performance of low-cadence calibration and subset redundant calibration using simulated visibilities and antenna gains. We use hexagonal array layouts, like the one shown in Figure~\ref{fig:hexarray}, with varying number of antennas for the simulations. Though this layout is loosely based on HERA (see \citet{dillon_parsons2016}), we expect the derived trends to hold for any two dimensional redundant array layout. We assume perfect redundancy in the array and identical antenna beams.

The simulations start by generating a set of true visibilities for all the unique baseline-types and a set of gains for all the antennas in the array. These simulated true visibilities have a random constant average amplitude across all baseline-types, are constant in time and uncorrelated between baseline-types. While this does not reflect a real sky signal, it is sufficient for the purpose of this paper because redundant-baseline calibration only has a weak dependence on the actual signal observed. The simulated antenna gains are Gaussian-distributed, with an average amplitude of 1. The antenna-to-antenna variation, or gain scatter, is assumed to be at the $\sim$10\% level which was found to be typical for HERA antennas~\citep{Kern_et_al_2020}.

The process of calibration in the FFT-correlator (first step in the yellow boxed region of Figure~\ref{fig:systemlayout}), is simulated by applying the estimated gains to the full visibility matrix. This is equivalent to applying the gains to individual antenna voltages and cross-correlating them. The full visibility matrix computed by an FX-correlator, ${\bf V^{full}}$, is generated by multiplying the simulated gains and true visibilities and adding Gaussian random noise. In the simulations where an explicit SNR for the visibilities is not mentioned, we have assumed an SNR of 10 even though a more favourable SNR is expected in practice. The visibilities produced by the FFT-correlator ${\bf V^{unique}}$ are generated by applying the gains estimated by one of the reduced redundant-baseline calibration processes to this full visibility matrix (Equation~\ref{eq:modelvis}).

For low-cadence calibration, the visibility matrix computed within the calibrator, ${\bf V^{reduced}}$, is generated by adding higher amplitude Gaussian noise to the multiplied gains and true visibilities. For subset redundant calibration, this visibility matrix is generated by choosing the visibilities of the selected antenna pairs from the full visibility matrix. Reduced redundant-baseline calibration is performed on this visibility matrix using the \texttt{omnical} algorithm, with a damping factor of 0.3 and convergence criteria of $10^{-10}$. The amplitude and phase degeneracies of the resulting gains are fixed by comparing with the amplitude and phase of the simulated input gains. Unless otherwise specified, the variance of antenna gains is computed by running 256 simulations with the same underlying gains and different realisations of true visibilities.

 
\section{Low-Cadence Calibration}
\label{sec:lowcadcal}

Low-cadence calibration is a reduced redundant-baseline calibration scheme that estimates antenna gains from visibilities that have been computed in a round-robin fashion. The calibrator in Figure~\ref{fig:systemlayout} cross-correlates the baselines required for calibration, but the computational resources allocated to it cannot scale faster than $\mathcal{O}(N\log{N})$. The computational resources required to compute visibilities is determined by the number of baselines that need to be cross-correlated \textit{simultaneously}.
By decreasing the number of antenna pairs that need to be correlated at a time, the computational resources required by the calibrator can be reduced. The full visibility matrix is populated after a few cycles and this is used to redundantly calibrate the array. Since redundant-baseline calibration can only be performed once in a given number of cross-correlation cycles, this calibration scheme is called low-cadence calibration.

The size of a low-cadence calibrator is determined by two parameters-- the time period available for generating the full visibility matrix within the calibrator ($t_{\rm{cal}}$) and the integration time allotted to each cycle ($t_{\rm{int}}$). These are related by the equation:
\begin{equation}
    t_{\rm{cal}} = N_{\rm{cycle}} \times t_{\rm{int}}
\end{equation}
\noindent
where $N_{\rm{cycle}}$ is the number of integration cycles taken by the calibrator to populate the full visibility matrix. The size of the low-cadence calibrator is inversely proportional to $N_{\rm{cycle}}$ i.e., for a small calibrator size we require a large $t_{\rm{cal}}$ and a small $t_{\rm{int}}$. 

Redundant-baseline calibration operates under the assumption that antenna gains and true visibilities are constant during the time period required to compute all the visibilities involved in the system of equations. For an FX-correlator, this is equal to the integration time of the full visibility matrix which is usually smaller than the inherent gain variability, and necessarily smaller than the time period over which the visibilities evolve due to a constantly rotating sky. For a low-cadence calibrator, the constancy of antenna gains and true visibilities within a calibration cycle has to be manually enforced.

The upper limit of $t_{\rm{cal}}$ is set by the inherent gain variability of the array, which could depend on numerous factors including the analog signal chain, the radio frequency interference environment and the precision of antenna gains required for the science application. If the time taken to generate the full visibility matrix is larger than the interval within which gains can be assumed to be constant, redundant-baseline calibration can result in erroneous gain solutions. If the time period of gain variability is large, it is possible that the true visibilities change within this period. However, to preserve redundancy we only require that all pairs of antennas with the same baseline be correlated simultaneously. Since this is necessarily always less than $N$ visibilities, a calibrator which can cross-correlate at least $N$ baselines can accommodate the largest redundant-baseline group in the array and satisfy this condition. 

A realistic lower limit for $t_{\rm{cal}}$ is the integration time of visibilities in the FFT-correlator. Within this period, the assumption of constant antenna gains and true visibilities holds and redundant-baseline calibration can be solved using the algorithms currently available. While $t_{\rm{cal}}$ can theoretically be set to a smaller value, it could unnecessarily increase the size of the calibrator by decreasing $N_{\rm{cycle}}$ for a given $t_{\rm{int}}$.

\subsection{Scaling in Gain Variance with Integration Time}

The relationship between integration time and SNR of a measured visibility is given by the radiometer equation \citep*[see][Appendix 1.1]{tms2017}. Substituting the radiometer equation into the variance of antenna gains in Equation~\ref{eq:gaincov} we get:

\begin{align}
    \label{eq:lowcadcal_gaincov}
        \sigma_{g}^2 &\approx (\rm{SNR})^{-2} \; \rm{diag}\left[\left(\matr{A}^{\dagger}\matr{A}\right)^{-1}_{(\mathit{N}\times \mathit{N})}\right] \nonumber \\
        &\propto \left(\sqrt{t_{\rm{int}}}\right)^{-2} \;
        \rm{diag}\left[\left(\matr{A}^{\dagger}\matr{A}\right)^{-1}_{(\mathit{N}\times \mathit{N})}\right] \; \propto \frac{1}{\mathit{t}_{\rm{int}}}
\end{align}
\noindent
which quantifies the variance of the antenna gains estimated by performing redundant-baseline calibration on visibilities integrated for a given duration. A shorter integration time leads to lower SNR in the measured visibilities and consequently, a higher variance in the antenna gains.

\begin{figure}
    \centering
    \includegraphics[width=\linewidth]{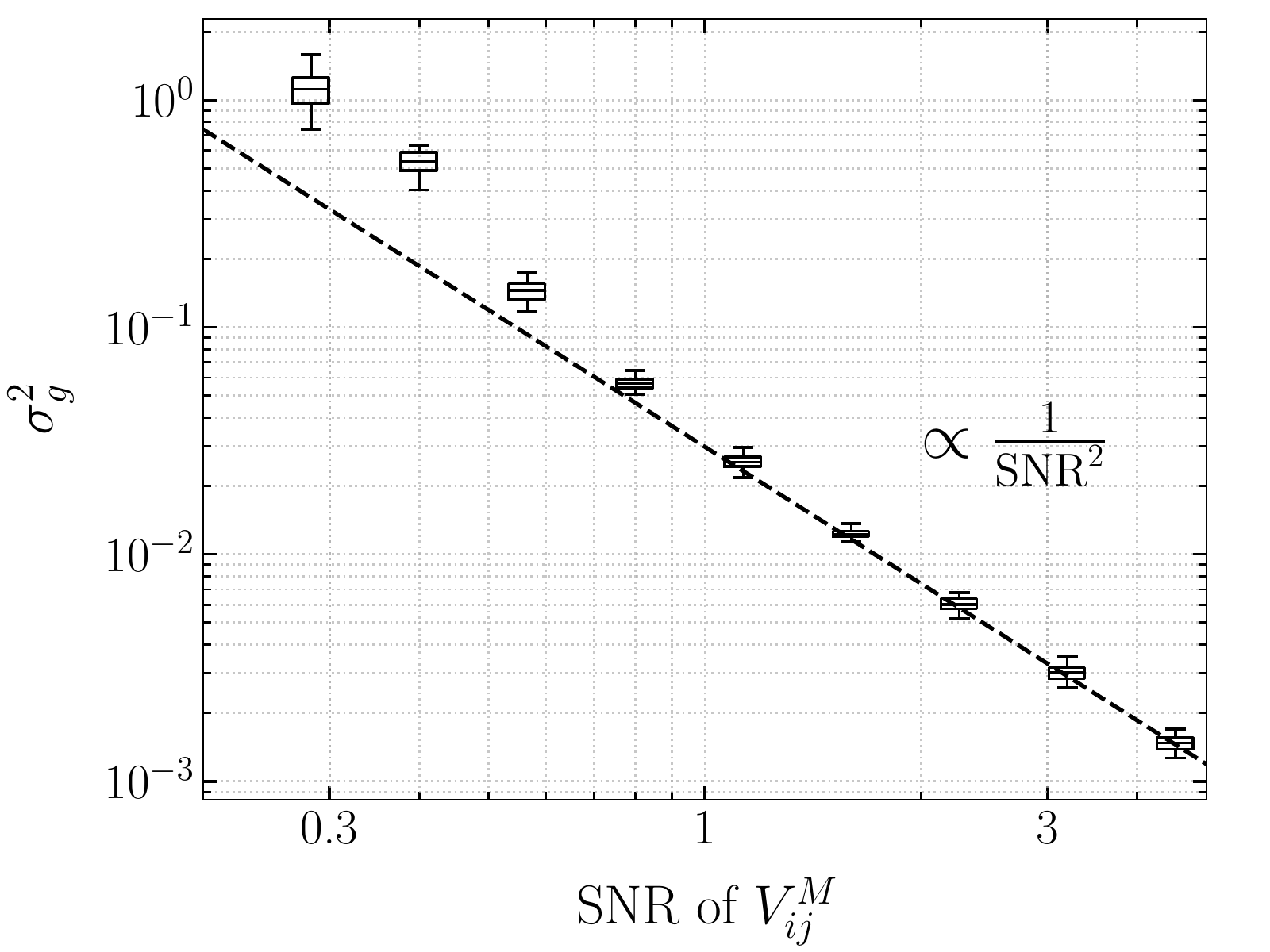}
    \caption{Relationship between the variance of antenna gains and the SNR of visibilities used for low-cadence calibration. The dashed line represents the inverse square relationship predicted by Equation~\ref{eq:gaincov} and the boxes show the results of simulation. The estimated variance of antenna gains in each simulation, is a distribution of $N$ points, where $N$ is the number of antennas in the array. The upper and lower limits of the `box' represent the upper and lower quartiles of the distribution. The horizontal bar within the box represents the mean of the distribution, and the whiskers show the total range of the estimated gain variances. At high SNR the gain variance follows the theoretically expected trend. At low SNR the variance is higher than the predicted value and becomes solver dependent because the $\chi^2$ is not effectively minimised by the solver.
    \label{fig:lowcadcal_wsnr}}
\end{figure}

Figure~\ref{fig:lowcadcal_wsnr} shows the trend in estimated gain variance with the average SNR of all the visibilities in an array. Each box in the figure represents the distribution of gain variance over all antennas in the array and has been generated assuming that all the baseline-types in the array have the same specified SNR. When the average SNR of visibilities is high, the estimated antenna gain variance follows the expected inverse square relationship. At low SNR, when the theoretical gain variance estimated using Equation~\ref{eq:gaincov} becomes comparable to the antenna-to-antenna scatter in gains squared ($\sim$0.01 for this simulation), redundant-baseline calibration fails at estimating antenna gains. Below a threshold SNR, that is set by the gain scatter, the inverse square relationship breaks down and the variance of estimated antenna gains becomes dependent on the solver. That is, the logarithmic approach, Taylor expansion approach, and \texttt{omnical} algorithms of linearizing the system of equations, result in different deviations from the given trend. This is because each algorithm minimises $\chi^2$ in a different way and none of them are effective at converging to the solution.

\begin{figure}
    \centering
    \includegraphics[width=\linewidth]{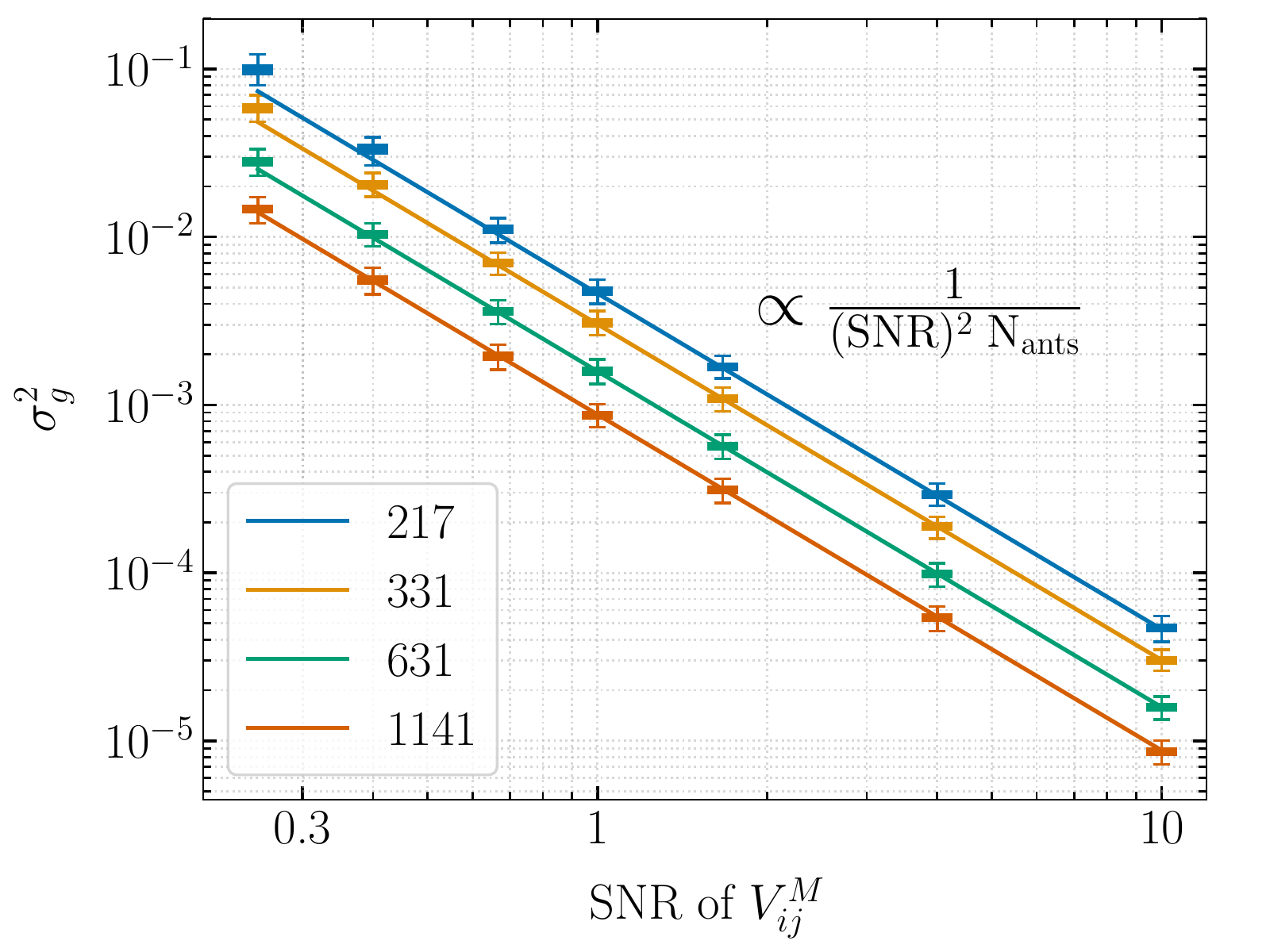}
    \caption{Relationship between distribution of gain variance and SNR of visibilities used in low-cadence calibration. The four different colours
    represent hexagonal arrays of various sizes; legend shows the number of 
    antennas in each layout. The solid lines represent the theoretical trend predicted by Equation~\ref{eq:gaincov} and the boxes show the distribution of gain variance obtained from simulation (see Figure~\ref{fig:lowcadcal_wsnr} for explanation). The threshold SNR, below which the variance deviates from the theoretical trend, is dependent on the number of antennas in the array.
    \label{fig:lowcadcal_wants}}
\end{figure}

\subsection{Scaling in Gain Variance with an $\mathcal{O}(N\log{N})$ Calibrator}

Figure~\ref{fig:lowcadcal_wants} shows the trend in gain variance with SNR of measured visibilities for hexagonal layouts with different number of antennas. The variance is suppressed by a factor of $N$ as the number of antennas in the array increase, because of the $\left(\matr{A}^{\dagger}\matr{A}\right)_{(\rm{N}\times \rm{N})}^{-1}$ term in Equation~\ref{eq:gaincov}. Notice that in larger arrays, the gain variance follows the theoretical trend even when the SNR of measured visibilities is less than one.

The threshold SNR below which the gain variance diverges from the theoretical prediction, changes with the number of antennas in the array. When the theoretical gain variance is less than the square of the expected gain scatter, the gain variance follows the inverse square relationship \textit{even when the SNR of measured visibilities is less than one}. This result is important because it allows low-cadence calibration to be scaled to extremely large arrays. 

Say, the computational resources allocated to the calibrator are restricted to scale similarly as the FFT-correlator. The calibrator cross-correlates $pN\log{N}$ baselines in each integration cycle, where $p$ is a pre-factor (like a proportionality constant) to convert the $\mathcal{O}(N\log{N})$ scaling into number of baselines. A larger pre-factor would result in a larger calibrator size. For a fixed interval of calibration $t_{\rm{cal}}$, the integration time is smaller for larger arrays according to the scaling: 

\begin{equation}
\label{eq:lowcadcal_inttime}
    t_{\rm{int}} = \left(\frac{pN\log{N}}{N^2/2}\right) t_{\rm{cal}} 
                 = \left(\frac{2p\log{N}}{N}\right) t_{\rm{cal}}
\end{equation}
\noindent
Hence, even if the interval of calibration is large, the multiplying factor might become small enough to push the SNR of measured visibilities to less than one. Substituting the result of Equation~\ref{eq:lowcadcal_inttime} into Equation~\ref{eq:lowcadcal_gaincov}, we get:

\begin{equation}
    \label{eq:gainvar_lowcadcal}
        \sigma_{g}^2 \propto \left(\frac{N}{2p\log{N}} \cdot  \frac{1}{t_{\rm{cal}}} \right) \frac{1}{N} \propto \left(\frac{1}{p\log{N}}\right) \cdot \frac{1}{t_{\rm{cal}}}
\end{equation}
\noindent
which shows that the antenna gain variance improves with array size even at constant $t_{\rm{cal}}$. That is, even though the SNR of measured visibilities might decrease to a value less than one, the antenna gain variance decreases. The price that one pays for not using $\mathcal{O}(N^2)$ resources for calibration is that the precision in antenna gains scales more slowly as compared to that of redundant-baseline calibration with the full visibility matrix which is given by Equation~\ref{eq:gainvar_redcal}.

Low-cadence calibration is a calculated way of trading computational resources for precision in the antenna gain solutions. As long as the size of the calibrator scales faster than $\mathcal{O}(N)$ with the size of the array, the variance in antenna gains decreases with increase in the number of antennas. A potential drawback of low-cadence calibration, especially when applied to arrays with over 10,000 antennas, is the time taken by a linearized solver to result in convergent gains. \citet{Dillon_et_al_2020} show that the time taken by the \texttt{omnical} algorithm scales as $\mathcal{O}(N^2)$ when the solver has to optimise $N^2$ baselines. If the time interval between calibration cycles $t_{\rm{cal}}$ can be proportionally decreased, it might still be possible to obtain real-time solutions. However, $t_{\rm{cal}}$ is usually set by the inherent gain variability in the array which might not be scalable with array size. One way of addressing this issue, is to look at redundant-baseline calibration with a limited set of baselines.

\section{Subset Redundant Calibration}
\label{sec:subredcal}

The spatial Fourier transform in an FFT-correlator averages visibilities of redundant baselines. Traditional redundant-baseline calibration assumes that all the $\sim$$N^2/2$ cross-correlation products are available for calibration, which could be non-viable to compute for large-N arrays. Subset redundant calibration is a reduced redundant-baseline calibration scheme that attempts to estimate antenna gains from visibilities of only a limited set of baseline-types. This section examines the effect of not using all baselines for redundant-baseline calibration on the variance of estimated antenna gains. 

While considering baselines for subset redundant calibration, it is useful to distinguish between baseline-types (or unique baselines) and redundant baseline groups. A baseline-type that a particular antenna pair belongs to is specified by the displacement vector between the two antennas. A redundant baseline group consists of all the antenna-pairs that have the same baseline vector. For instance, Figure~\ref{fig:hexarray} shows three different baseline-types with displacement vectors pointing East (grey), Southeast (blue) and Southwest (orange). Each baseline-type has 30 different antenna pairs in its redundant baseline group, marked in arrows of the same colour. 

\subsection{Brief Discussion on Using Short Baselines}

The short baseline-types are important in subset redundant calibration for two main reasons: (a) they involve all the antennas in the array, allowing redundant-baseline calibration on just these visibilities to estimate all the antenna gains and (b) the redundant baseline groups of these baseline-types are larger than the longer baseline-types. For example, in the layout shown in Figure~\ref{fig:hexarray}, there are 30 baselines in each of the redundant baseline groups that belong to the shortest three baseline-types, while there are only 4 baselines that belong to the group formed by the baseline-type like $(0,31)$. This is important for subset redundant calibration because every new baseline-type added to the system of equations requires a new variable in the form of the unique visibility for that baseline-type. Since the short baselines have a higher ratio of redundant baselines (measurements) to unique visibilities (variables), they contribute more to constraining the gain solutions. 

In addition to this, at low radio frequencies, the short baselines pick up the bright diffuse emission from our galaxy and have high SNR visibilities. As shown in Equation~\ref{eq:gaincov}, this suppresses gain variance and results in higher precision gain solutions. \citet{orosz_et_al2018} discuss other advantages of using only short baselines from the perspective of non-redundancies in a realistic array layout. They argue that calibration errors affect the inferred power spectrum, and contamination worsens when longer baselines are included in the redundant-baseline calibration process. \citet{li_et_al2018} point out that redundant-baseline calibration performs better than sky-based calibration at low radio frequencies, partly because short baselines have to be ignored for sky-based calibration due to poor diffuse sky models. On the other hand, the shorter baselines are more susceptible to systematic errors like antenna cross-coupling~\citep{Kern_et_al_2019} and may be more non-redundant than the longer baselines \citep{Dillon_et_al_2020} in a realistic array layout. 

A practical subset redundant calibrator would cross-correlate a combination of short and long baselines that produces the best estimate of antenna gains for the array. Since the voltages from all the antennas are available to the subset redundant calibrator, the combination of baselines that it needs to compute can also dynamically change with time/day of observation. In this paper, the baselines used to perform subset redundant calibration are considered in the order of baseline length from shortest to longest. That is, a smaller calibrator preferentially cross-correlates only the shorter baselines. However, the results presented in this section apply to combinations of short and long baselines as well.

\subsection{Degeneracy Criterion}

When using a limited set of baselines for redundant-baseline calibration, it is important to ensure that there are sufficient number of measurements to determine all the variables. The solution space of Equation~\ref{eq:redcal} has a null space with four degenerate parameters \citep{wieringa1992, liu_et_al2010, dillon_et_al2018} --  
\begin{enumerate*}[label={(\alph*)}]
\item the absolute amplitude of the gains (or the sum of all gains), 
\item the absolute phase of the gains (or the sum of all gain phases), 
\item the phase slope of the gains in the $x$ direction and 
\item the phase slope of the gains in the $y$ direction. 
\end{enumerate*} 

When selecting baselines for subset redundant calibration, it is important to verify that the null space of the solution set is restricted to these four degenerate parameters. Introducing more degeneracies allows the gain solutions to vary in that dimension and could require future corrections. If the additional degeneracy does not have a physical interpretation this may not even be possible. For a hexagonal layout like the one shown in Figure~\ref{fig:hexarray}, a minimum of three unique baseline-types, with displacement vectors in the directions marked in the figure, are required to satisfy the degeneracy requirement.

\subsection{Scaling in Gain Variance with Number of Baselines}

\begin{figure}
    \centering
    \includegraphics[width=\linewidth]{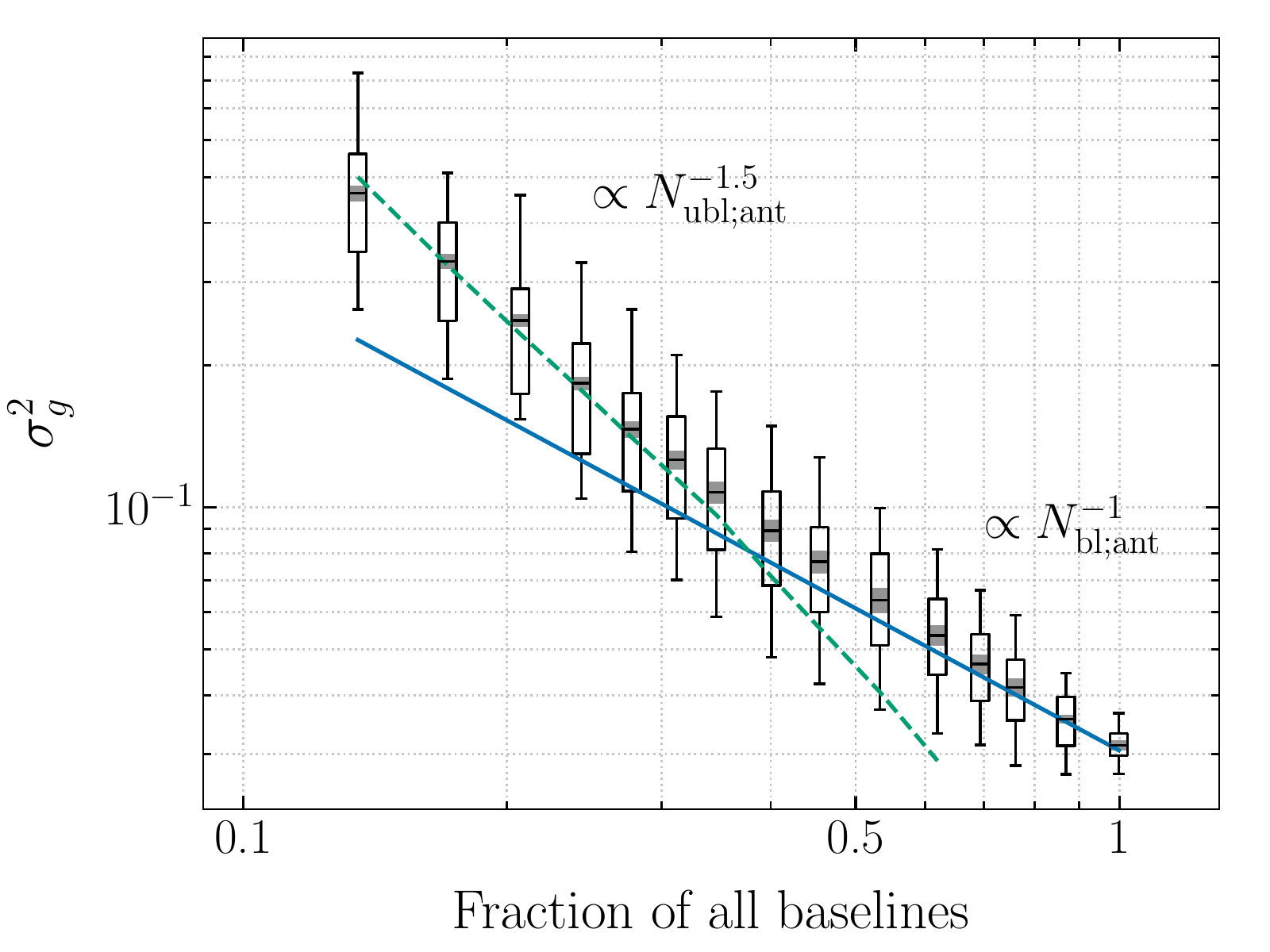}
    \caption{Relationship between fraction of baselines used in subset redundant calibration and the resulting variance in estimated gains. The boxes show the distribution of gain variances in simulation (see Figure~\ref{fig:lowcadcal_wsnr} for explanation), where increasingly longer baselines are included in the subset redundant calibration system of the simulation. When the fraction of baselines used in estimating antenna gains is small, the gain variance depends on the number of unique visibilities per antenna used in calibration ($N_{\rm{ubl;ant}}$); this trend is shown by the dashed green line. When a large fraction of baselines are used for calibration, the gain variance depends on the total number of measurements per antenna ($N_{\rm{bl;ant}}$); shown by the solid blue line. A second set of boxes in grey show the antenna-to-antenna variation with all the edge antennas excluded.
    \label{fig:subredcal_wbls}}
\end{figure}

Figure~\ref{fig:subredcal_wbls} shows the relationship between number of baselines used in subset redundant calibration and the corresponding variance in the estimated gains. The number of baselines are shown in terms of a fraction of the total baselines in the array. The black boxes show the results of simulation (see Section~\ref{sec:simulation}). The x-axis from left to right, represents baselines added to the subset redundant calibration system in ascending order of baseline length (starting with the minimum required to satisfy the degeneracy criterion). For each baseline-type added to the system, it is assumed that all the redundant baseline pairs that contribute to that baseline-type are used for calibration. The SNR of visibilities is assumed to be similar for all baseline-types (unlike for a real sky), and constant through the simulation. The vertical range of the boxes, which represents distribution of gain variance over antennas in the array, is larger than the case of low-cadence calibration. However, most of this variation comes from the edge antennas in the array. A second set of solid boxes in grey show the distribution of gain variances with the edge antennas excluded. The reason for this decrease in variation is discussed in more detail in Section~\ref{sec:subredcal_cov}.

The solid line, in blue, represents the inverse of the total number of baselines per antenna ($N_{\rm{bl;ant}}$) that are used to perform subset redundant calibration.

\begin{equation}
    \label{eq:subredcal_n_bl}
    N_{\rm{bl;ant}} = \frac{N_{\rm{obs}}}{N} \approx \frac{f \; N^2/2}{N} = f \; \frac{N}{2}
\end{equation}
\noindent
Here $f$ represents the fraction of all baselines used in subset redundant calibration. When the fraction of baselines used in subset redundant calibration is high, the gain variance asymptotes to the inverse of the total number of measurements-per-antenna in the system of equations. This trend is expected from the Gaussian noise in visibility measurements; each new measurement added to the system contributes to decreasing the noise in the estimated gains. 

When the fraction of baselines used in estimating antenna gains is small, gain variance depends on two factors: (a) the number of baseline-types included in subset redundant calibration, and (b) the number of antennas that are involved in forming redundant baselines for these baseline-types. A combination of these two factors is captured by the variable $N_{\rm{ubl;ant}}$ which represents the average number of unique visibilities per antenna.

\begin{equation}
    \label{eq:subredcal_n_ubl}
    N_{\rm{ubl;ant}} = \frac{1}{N} \sum\limits_{\alpha} N_{\rm{ant}} \left[\in V_{\alpha} \right]
\end{equation}
\noindent
In the above equation, the variable being summed is the number of antennas that are involved in forming redundant baselines of the baseline-type $V_{\alpha}$. The summation runs over all the baseline-types that are used in subset redundant calibration and $N$ is the total number of antennas in the array. Consider the case where all the baseline-types used in subset redundant calibration have redundant baselines involving all the antennas in the array (for instance, when only the shortest 3-6 baseline-types are used for calibration). In such a system, $N_{\rm{ubl;ant}}$ simply evaluates to the total number of baseline-types (or unique visibilities) used in the calibration process. If some of the baseline-types used in subset redundant calibration involve only a couple of antennas, $N_{\rm{ubl;ant}}$ is smaller than the total number of unique visibilities in the system of equations. 

Empirically, we find that the relationship between gain variance and the average number of unique visibilities per antenna is a power law with a slope around $-1.5$ for hexagonal and square layouts. This power law trend is shown by the dashed green line in Figure~\ref{fig:subredcal_wbls}. The large antenna-to-antenna variation in this regime makes it difficult to determine the exact slope or understand the origin of this power law. We suspect that it originates in the way gain error propagates from antenna to antenna. 

In summary, for a subset redundant calibration system that uses only a small fraction of the total baselines in the array, gain variance improves when baseline-types with larger redundant groups are used for calibration.

\subsection{Covariance in Estimated Gains}
\label{sec:subredcal_cov}

The improvement in gain variance obtained when using a higher number of unique baselines per antenna, can also be explained through the gain covariance. When the fraction of baselines used in calibration is small, in addition to high variance, the gains also have a relatively high covariance. The covariance in gains is given by the off-diagonal terms of the marginalised covariance matrix:

\begin{equation}
    \label{eq:gaincov_v1}
    \matr{C^{\prime}} \approx \left(\rm{SNR}\right)^{-2} \left(\matr{A}^{\dagger}\matr{A}\right)_{(\mathit{N} \times \mathit{N})}^{-1}
\end{equation}
\noindent
Both the diagonal and off-diagonal terms of the matrix $\left(\matr{A}^{\dagger}\matr{A}\right)$ change when a subset of baseline-types are used for redundant calibration. Inverting this matrix changes the covariance in the resulting gains.

\begin{figure}
    \centering
    \includegraphics[width=\linewidth]{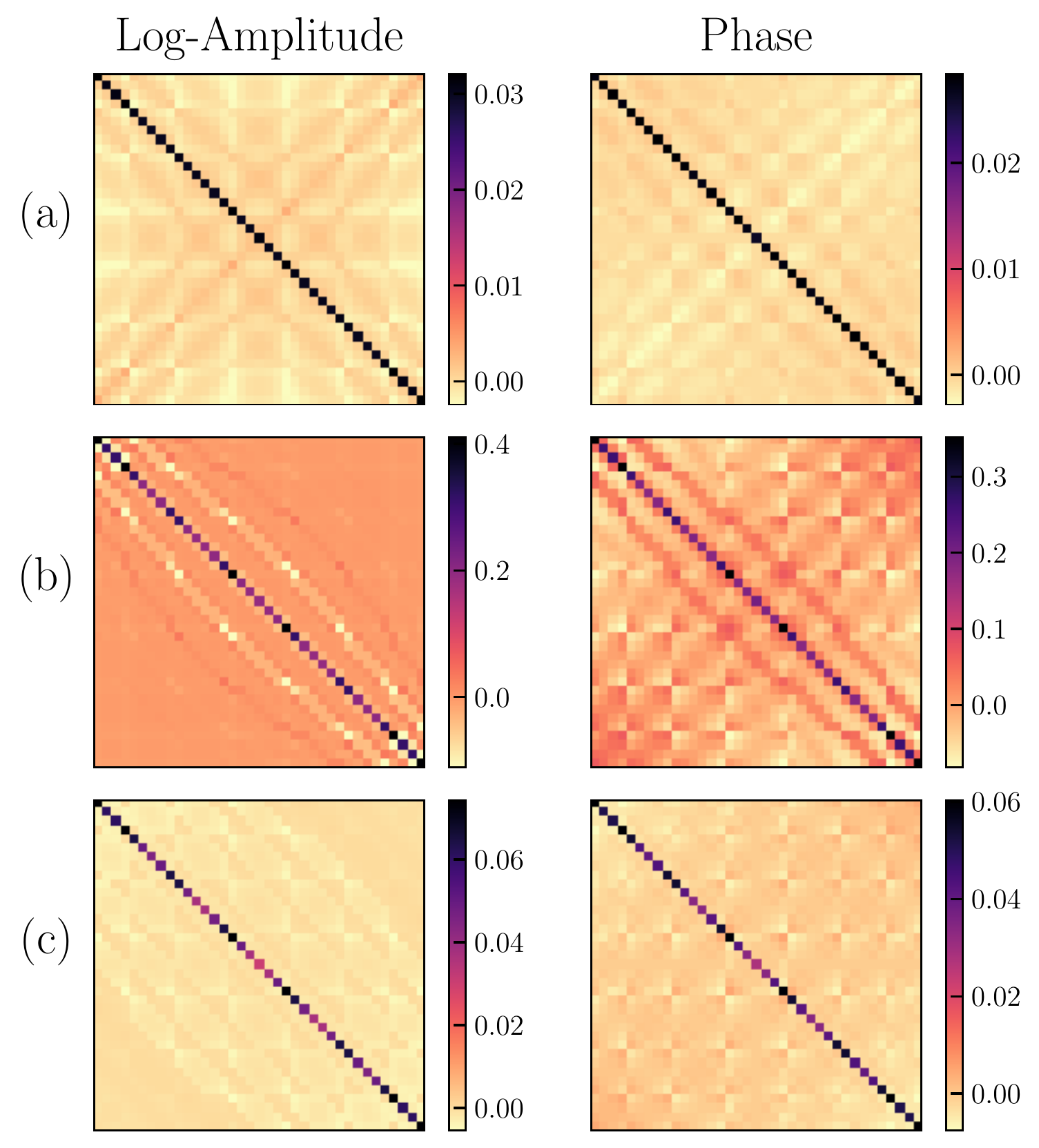}
    \caption{Covariance in the amplitude and phase of antenna gains for subset redundant calibration performed on the hexagonal layout of Figure~\ref{fig:hexarray}. Panel (a) shows the marginalised covariance for the case where redundant-baseline calibration is performed using the full visibility matrix. Panel (b) shows the covariance matrix when subset redundant calibration is performed with just the shortest three baselines required to satisfy the degeneracy criterion. In addition to higher gain variance, the antennas have also have a high covariance. Panel (c) is the covariance matrix for subset redundant calibration performed using more than half of the total baselines in the array. Both the variance and the covariance are comparable to that of redundant-baseline calibration performed with the full visibility matrix even though the covariance has a different structure compared to panel (a), and the variance is still dependent on antenna location.
    \label{fig:subredcal_cov}}
\end{figure}

Figure~\ref{fig:subredcal_cov} shows the marginalised covariance matrices for three different subset redundant calibration systems (for the hexagonal array layout in Figure~\ref{fig:hexarray}). The logarithmic approach to linearizing, shown in Equation~\ref{eq:logcal}, naturally results in separating the amplitude and phase of gains into the real and imaginary parts of the logarithm respectively. Hence, using the real (imaginary) part of the matrix $\matr{A}$ gives the gain covariance in the amplitude (phase) of gains. All the covariance matrices are normalised by the thermal noise in the visibilities used for redundant-baseline calibration. 

Panel (a) of Figure~\ref{fig:subredcal_cov} shows the covariance matrices for the case where redundant-baseline calibration is performed using the full visibility matrix. The variance in antenna gains, given by the diagonal of the matrix, has an average value of $1/N$ as predicted by Equation~\ref{eq:gainvar_redcal}. Though not evident in the figure, this variance is weakly dependent on antenna location as shown by \citet{dillon_parsons2016}. There a small but non-zero covariance in the antenna gains. 

Panel (b) of Figure~\ref{fig:subredcal_cov} shows the covariance matrices for the extreme case where only the shortest three baselines (the minimum baseline-types required to satisfy the degeneracy criterion) are used in redundant-baseline calibration. Notice that the variance is nearly an order of magnitude higher than the first case and clearly dependent on antenna location. The covariance between antennas is non-negligible and higher between antennas that have a high variance. 

Panel (c) of Figure~\ref{fig:subredcal_cov} shows the covariance matrices for the case where more than half the baselines are used in subset redundant calibration. This is the threshold at which gain variance starts following the inverse measurements per antenna trend in Figure~\ref{fig:subredcal_wbls}. Even though the variance is still antenna location dependent and the covariance has a different structure, they are comparable to the case of redundant-baseline calibration with the full visibility matrix. 

When the subset redundant calibration system involves a larger number of unique baselines, there are more independent constraints on the gain of each antenna. This decreases the average gain variance, covariance between antennas and also the antenna-location dependence of the variance. In Figure~\ref{fig:subredcal_wbls}, the errorbars associated with the gain variance estimated in simulation, represent the antenna-to-antenna variation within the simulation. Hence, the errorbars are larger for the gains estimated using a smaller fraction of baselines.

\begin{figure}
    \centering
    \includegraphics[width=\linewidth]{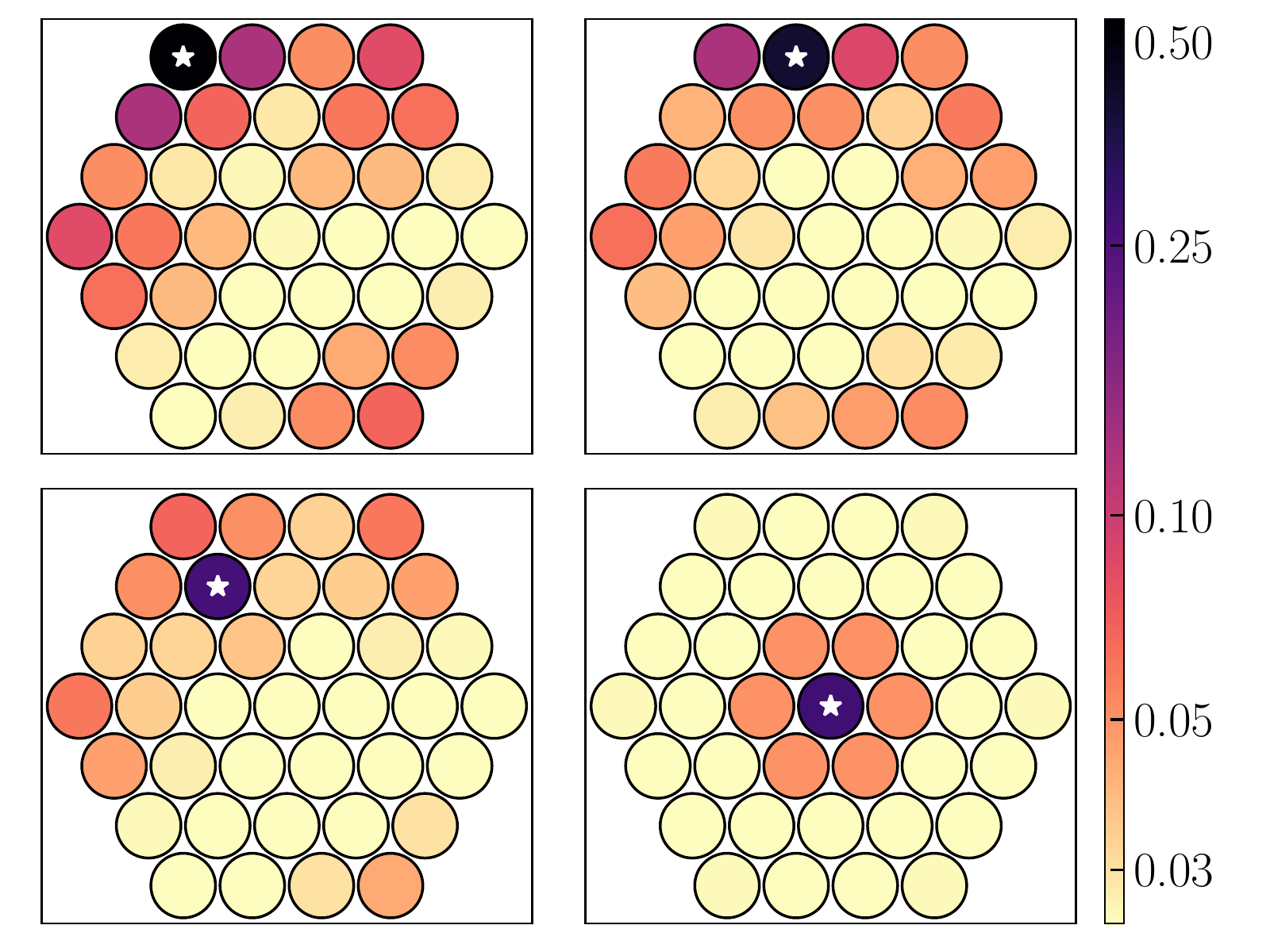}
    \caption{Four rows of the covariance matrix in Panel (b) of Figure~\ref{fig:subredcal_cov}. The colour-bar has been log-normalised to make the covariance more evident. The covariance shown in each panel is the sum of covariance in amplitude and phase for the antenna that is marked with a star. For all antennas, the covariance is highest with adjacent antennas because only those visibility measurements are used for constraining gain solutions. The variance and covariance of corner and edge antennas (upper two panels) are higher than that of antennas placed centrally (lower two panels) because the edge antennas participate in fewer visibility measurements than central antennas.
    \label{fig:subredcal_gaincov}}
\end{figure}

This dependence of antenna gain variance and covariance on the number of independent constraints per antenna is more evident in Figure~\ref{fig:subredcal_gaincov} which shows four different rows of the covariance matrix in Panel (b) of Figure~\ref{fig:subredcal_cov} in an exaggerated manner. In this case, subset redundant calibration is performed with just the shortest three baseline-types required to satisfy the degeneracy criterion. The covariance and variance of edge antennas is higher than that of centrally placed antennas because the corner antennas participate in fewer cross-correlations (three for the antenna in the top left panel, four for the antenna in the top right) than centrally placed antennas (six each for the antennas in the bottom two panels). Hence, there are fewer independent constraints for the edge antennas leading to a higher variance in their estimated gains. Ignoring the edge antennas after subset redundant calibration leads to an 80-30\% decrease in the antenna-to-antenna variation depending on the fraction of baselines used in the calibration process. The distribution of antenna gain variances with the edge antennas excluded is shown using solid grey boxes in Figure~\ref{fig:subredcal_wbls}.

\subsection{Scaling in Gain Variance with an $\mathcal{O}(N\log{N})$ Calibrator}

\begin{figure}
    \centering
    \includegraphics[width=\linewidth]{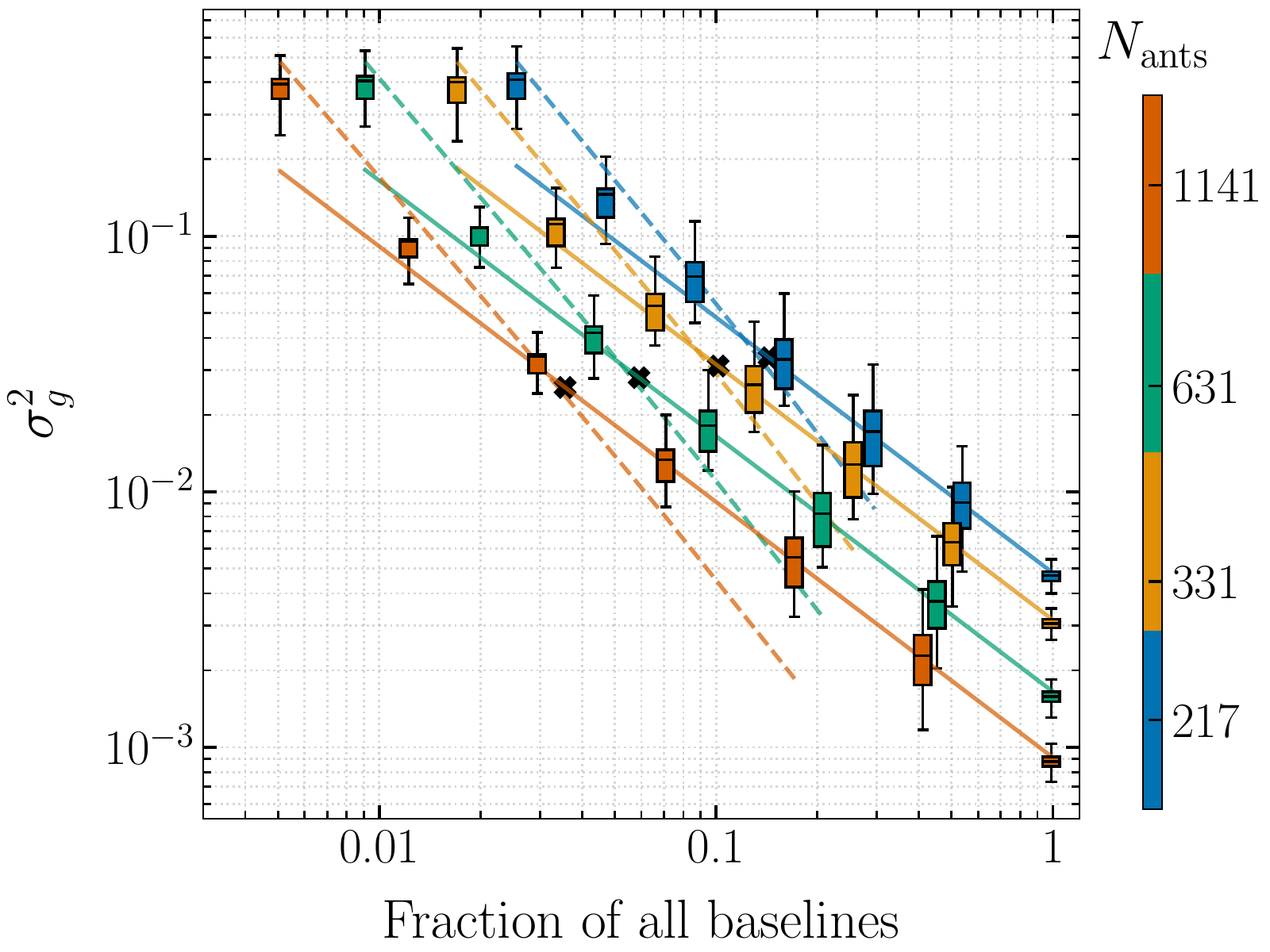}
    \caption{Relationship between variance of antenna gains and the fraction of baselines used in subset redundant calibration, for hexagonal array layouts of various sizes (shown in different colours). The boxes represent the distribution of gain variance obtained from simulation (see Figure~\ref{fig:lowcadcal_wsnr} for explanation). The solid lines represent an inverse trend of baselines per antenna $\left(N_{\rm{bl;ant}}^{-1}\right)$. The dashed lines show a power law dependence on the number of different baseline-types per antenna $\left(N_{\rm{ubl;ant}}^{-1.5}\right)$ used for calibration. The small black crosses represent the fraction of baselines that can be cross-correlated by a calibrator that can process $N\log{N}$ baselines.
    \label{fig:subredcal_wants}}
\end{figure}

Figure~\ref{fig:subredcal_wants} extends the relationship between gain variance and fraction of baselines used in subset redundant calibration to hexagonal layouts with larger number of antennas. The four different colours represent four different array sizes. Within each colour, the boxes represent the distribution of gain variance over antennas in the array (see Section~\ref{sec:simulation}). The solid lines represent an the inverse measurements per antenna trend $\left(N_{\rm{bl;ant}}^{-1}\right)$ and the dashed lines show a power law relationship between gain variance and number of unique visibilities per antenna $\left(N_{\rm{ubl;ant}}^{-1.5}\right)$. It is evident that the two asymptotes to the gain variance, shown in Figure~\ref{fig:subredcal_wbls}, hold with changing array size. 

When a large fraction of baselines are used for subset redundant calibration, the gain variance depends on the total number of baselines used for calibration. At a constant fraction of baselines, the number of baselines formed by an antenna is proportional to the number of antennas in the array as shown in Equation~\ref{eq:subredcal_n_bl}. The four solid lines show a scaling in gain variance by this factor. When the fraction of baselines used for calibration is small, the gain variance does not necessarily decrease with increase in array size. The leftmost point within each array size represents the variance in the estimated gains when subset redundant calibration is performed using just the shortest three baseline-types (required for the degeneracy criterion). As evident from the figure, the gain variance is nearly constant despite the larger number of redundant baselines in an array with more antennas. This is because, within increasing array size, there are as many new variables (in the form of antenna gains) as measurements. For improvement in variance, the size of the subset redundant calibrator should increase with the array size.

Say the computational resources allocated to a subset redundant calibrator are restricted to scale similarly to the FFT-correlator. The calibrator can process $pN\log{N}$ baselines for a given array size. The fraction of baselines that can be processed by such a calibrator is given by:

\begin{equation}
    \label{eq:subredcal_frac_bls}
    f = \frac{pN \log{N}}{N^2/2} = \frac{2p\log{N}}{N}
\end{equation}
\noindent
Figure~\ref{fig:subredcal_wants} shows this fraction of baselines, for a pre-factor of one, in small black crosses. When the calibrator cross-correlates exactly $N\log{N}$ baselines, this fraction falls in the transition region between the two asymptotes. However, assuming that the $N_{\rm{bl;ant}}^{-1}$ approximation holds at this fraction of baselines, we can substitute Equation~\ref{eq:subredcal_frac_bls} into the number of baselines per antenna in Equation~\ref{eq:subredcal_n_bl} to get the overall scaling in gain variance.

\begin{align}
    \label{eq:gainvar_subredcal}
        \sigma_{g}^2 &\propto N_{\rm{bl;ant}}^{-1} \nonumber \\
        &\propto \displaystyle \frac{1}{f \; \frac{N}{2}} \propto \frac{1}{p\log{N}}
\end{align}
\noindent
The scaling in gain variance with array size, using subset redundant calibration, is similar to that obtained using low-cadence calibration (Equation~\ref{eq:gainvar_lowcadcal}). In both cases, the price one pays for not using $\mathcal{O}(N^2)$ baselines for calibration is that the gain variance scales slower than the case where redundant calibration is performed using the full visibility matrix (Equation~\ref{eq:gainvar_redcal}).

Subset redundant calibration leverages the higher constraining power of some baseline-types, by allocating computational resources of the calibrator to preferentially cross-correlating those redundant baselines. The least number of baselines that need to be considered is set by the null space of the solution to the redundant-baseline calibration equations. However, using a small fraction of baselines can result in antenna gains that have a non-negligible covariance and location-dependent variances. If the fraction of baselines cross-correlated by a subset redundant calibrator can scale as $\mathcal{O}(N\log{N})$ or faster, the gain variance decreases with the increase in array size and scaling in gain variance is similar to that of low-cadence calibration.

\section{Comparison between the Performance of Low-Cadence Calibration and Subset Redundant Calibration}
\label{sec:comparison}

Low-cadence calibration and subset redundant calibration are two potential solutions to calibrating FFT-correlators for redundant array layouts without computing $\mathcal{O}(N^2)$ cross-correlations. In low-cadence calibration, the calibrator computes the visibilities of all baselines through a round-robin of antenna pairs and spends a shorter amount of time on each visibility measurement. As a consequence, the SNR of measured visibilities is lower and leads to a higher variance in the estimated gains. In subset redundant calibration, the calibrator computes the correlations of only a few baselines (preferentially the shorter baseline pairs) and uses these visibilities to estimate the antenna gains. In this case, having fewer measurements leads to higher variance in the estimated gains.

\subsection{Scaling in Gain Variance}

\begin{figure}
    \centering
    \includegraphics[width=\linewidth]{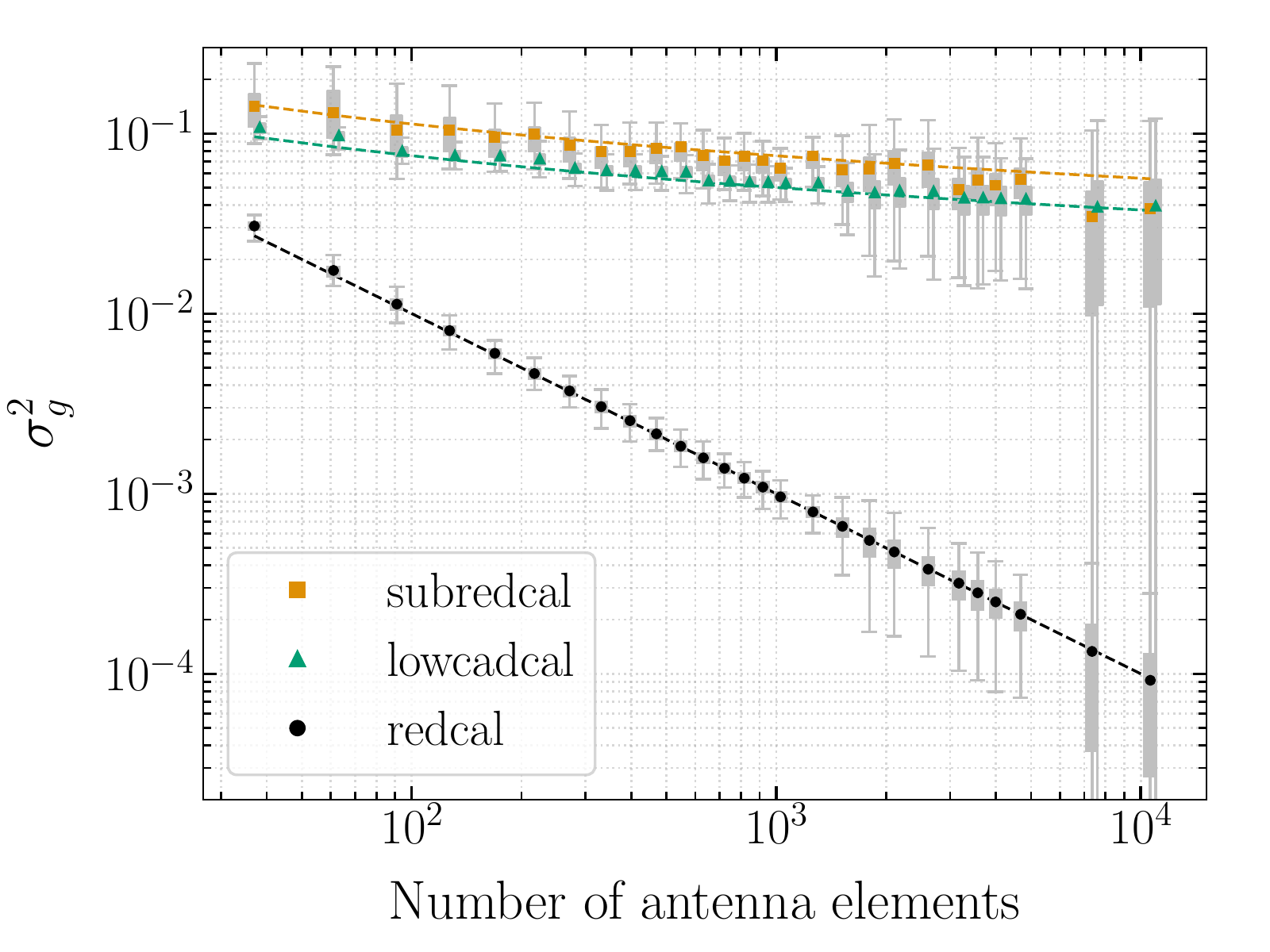}
    \caption{Distributions of antenna gain variance, estimated using low-cadence calibration (\texttt{lowcadcal}), subset redundant calibration (\texttt{subredcal}) and traditional redundant-baseline calibration (\texttt{redcal}) for hexagonal layouts of various sizes. The markers show the mean of the distribution of gain variance for each array size, and the distribution itself is shown using grey solid boxes around each marker (see Figure~\ref{fig:lowcadcal_wsnr} for explanation). The larger vertical extent of boxes for arrays with $N>1500$ elements are due to fewer simulations. The dashed lines represent the theoretically derived trends (Equations~\ref{eq:gainvar_redcal},~\ref{eq:gainvar_lowcadcal},~\ref{eq:gainvar_subredcal}). Low-cadence calibration consistently yields lower gain variance than subset redundant calibration for similar calibrator sizes. While both the reduced redundant-baseline calibration techniques cross-correlate $\sim$$N\log{N}$ baselines for each array size, traditional redundant-baseline calibration assumes that the complete visibility matrix with $\sim$$N^2/2$ baselines is available for calibration. The additional reduction in gain variance for the latter comes at a high computational cost and may not be necessary when scatter in redundant visibilities is dominated by thermal noise (see Equations~\ref{eq:visvar_lowcadcal} and~\ref{eq:visvar_subredcal}).
    \label{fig:scale_ants}}
\end{figure}

Figure~\ref{fig:scale_ants} compares the scaling in the average variance of gains estimated using either reduced redundant-baseline calibration method to the scaling in the average variance of gains estimated using redundant-baseline calibration on the full visibility matrix measured at high SNR. For both low-cadence calibration and subset redundant calibration, we assume that the pre-factor $p=1$ and use only $N\log{N}$ visibility measurements to estimate antenna gains. The distribution of gain variances, for arrays with number of antennas in the range 1500-5000, is computed using only 16 simulations (instead of 256) due to long simulation times. For the same reason, the distributions for the two largest array sizes with $N>5000$ antennas are computed using only 2 simulations each.

For low-cadence calibration, the interval between calibration cycles, $t_{\rm{cal}}$ is assumed to be constant for all array sizes. The integration time $t_{\rm{int}}$ has been scaled down according to Equation~\ref{eq:lowcadcal_inttime} to keep the size of the calibrator at an $\mathcal{O}(N\log{N})$ scaling. As predicted, the variance in the estimated antenna gains scales according to Equation~\ref{eq:gainvar_lowcadcal} and is shown by the dashed green line in Figure~\ref{fig:scale_ants}.

For subset redundant calibration, the SNR of measured visibilities is assumed to be constant for all array sizes. The number of baselines used in the calibration process for a given array size is the closest whole number, to the fraction given by Equation~\ref{eq:subredcal_frac_bls}, that accounts for an integer number of redundant baseline groups in ascending order of their baseline length. That is, baseline-types are considered in the order of their baseline length and added to the subset redundant calibration system only if all the redundant baselines that contribute to that baseline-type can be considered. This causes a non-uniform increase in the number of baseline-types used for calibration as the array size increases. The overall trend in gain variance follows the scaling predicted by Equation~\ref{eq:gainvar_subredcal} and is shown by the dashed orange line in Figure~\ref{fig:scale_ants}. 

Subset redundant calibration results in a higher gain variance than low-cadence calibration because, as shown in Figure~\ref{fig:subredcal_wants}, the approximation of inverse measurements per antenna (Equation~\ref{eq:gainvar_subredcal}) underestimates the gain variance when the pre-factor is unity. However, the scaling predicted by that approximation holds true for all the simulated array sizes. For two largest array sizes in Figure~\ref{fig:scale_ants}, the estimated gain variance using subset redundant calibration seems comparable to that of low-cadence calibration but is not an exception to this trend. This is because the two simulations used to compute these estimates generated a favourable set of visibilities for subset redundant calibration with the short baselines having larger amplitudes than the long baselines. In general, a subset redundant calibrator would have to cross-correlate more baselines than a low-cadence calibrator to achieve the same gain variance.

Redundant-baseline calibration on the full visibility matrix, measured at high SNR results in a $1/N$ scaling as predicted by Equation~\ref{eq:gainvar_redcal}. and shown by the black dashed line in Figure~\ref{fig:scale_ants}. Such a calibration method requires a calibrator with $\mathcal{O}(N^2)$ computational resources that may not be viable for large arrays. Moreover, the high precision in gains obtained using the full visibility matrix might not be necessary for decreasing the scatter in calibrated redundant visibilities.

\subsection{Variance in Calibrated Redundant Visibilities}

The visibilities of redundant baselines, that have been calibrated using a reduced redundant-baseline calibration processes, are averaged by the spatial Fourier transform. This converts any post-calibration residual scatter in redundant visibilities into additional noise on the unique visibilities returned by the FFT-correlator.

If the estimated antenna gains are exactly the true gains, the scatter in redundant-baseline averaged visibilities is given by $\sigma^2/N_{\alpha}$, assuming that the variance in the thermal noise of visibilities is similar for all baselines and represented by $\sigma^2$. However, gains estimated using redundant-baseline calibration diverge from the true gains with an average scatter that is given by Equation~\ref{eq:gaincov}. Hence, the calibrated redundant visibilities have a residual scatter that comes from both the thermal noise in the measurements and the variance in the estimated gains. 

The variance in calibrated visibilities can be derived using Equation~\ref{eq:modelvis} and the first order approximation for the variance of non-linear functions. In this derivation, we have assumed that the calibrated visibilities are not correlated with each other. Moreover, since the gains estimated using reduced redundant-baseline calibration are applied to a different set of visibilities that those used to estimate them, we can also ignore the covariance between visibilities and gains. The multiplying antenna gains, however, have a non-negligible covariance that is represented by the terms $\rho_{g_ig_j}$ in the following equation. This represents the off-diagonal components in the covariance matrix $\matr{C^{\prime}}$ in Equation~\ref{eq:gaincov_v1}.

\begin{equation}
    \sigma_{V_{\alpha}^{\rm{unique}}}^2 \approx \frac{\left|V_{\alpha}^{\rm{unique}}\right|^2}{N_{\alpha}^2} \;
    \sum\limits_{(i,j) \in \alpha} \left[\frac{\sigma_{ij}^2}{\left|V_{ij}\right|^2} + \frac{\sigma_{g_i}^2}{\left|g_i\right|^2} + \frac{\sigma_{g_j}^2}{\left|g_j\right|^2} + 2\frac{\rho_{g_i g_j}}{\left|g_ig_j\right|}\right]
\end{equation}
\noindent
We can simplify this further under the assumptions that the average amplitude of all antenna gains is close to one ($\left|g_i\right|^2$$\sim$1) and that the gain variance of all antennas is similar and given by $\sigma_g^2$. Note that the relative variance of visibilities is just the inverse squared SNR. In the equation below, $\rm{SNR}_{\alpha; \; \rm{unique}}$ is the SNR of the visibility computed by the FFT-correlator for the baseline-type represented by $\alpha$ and $\rm{SNR}_{\rm{full}}$ is the average SNR of the $N_{\alpha}$ number of redundant visibilities that belong to that baseline-type.

\begin{equation}
    \label{eq:visvar}
    (\rm{SNR})_{\alpha; \; \rm{unique}}^{-2} \approx \frac{1}{\mathit{N}_{\alpha}} \left[(\rm{SNR})_{\rm{full}}^{-2} + 2\sigma_\mathit{g}^2 + \frac{2}{\mathit{N}_{\alpha}} \sum\limits_{\mathit{(i,j)} \in \alpha} \rho_{\mathit{g}_\mathit{i}\mathit{g}_\mathit{j}}\right]
\end{equation}
\noindent
This equation gives the acceptable range of gain variance and covariance for the gains estimated using redundant-baseline calibration. When the gain variance and covariance is much smaller than the thermal noise in visibilities, the first term dominates the residual scatter. If this is satisfied, lowering gain variance by using a larger calibrator will not improve the variance in calibrated visibilities. 

\subsubsection{Variance in Redundant Visibilities Calibrated using Low-cadence Calibration}

The variance in antenna gains estimated using low-cadence calibration depends on the SNR of the visibilities computed in the calibrator which in turn depends on the integration time (Equation~\ref{eq:lowcadcal_gaincov}) available for each cycle of computation. The relationship between the SNR of the full visibility matrix and that of the reduced visibility matrix computed in by calibrator, that scales as $\mathcal{O}(N\log{N})$, can be written using Equation~\ref{eq:lowcadcal_inttime} as:

\begin{equation}
    \label{eq:SNR_lowcadcal}
    \displaystyle \frac{\left(\rm{SNR}\right)^{-2}_{\rm{full}}}{\left(\rm{SNR}\right)^{-2}_{\rm{reduced}}} = \frac{t_{\rm{int;reduced}}}{t_{\rm{int;full}}} = \left(\frac{2p\log{N}}{N}\right)\; \frac{t_{\rm{cal}}}{t_{\rm{int;full}}}
\end{equation}
\noindent
where $t_{\rm{int;full}}$ is the integration time used in an FX- or FFT-correlator. Substituting the above relation into Equation~\ref{eq:gaincov}, we can write the variance in estimated gains in terms of the SNR of the visibility matrix that the gains calibrate. 

\begin{align}
    \sigma_g^2 &= (\rm{SNR})_{\rm{reduced}}^{-2} \; \left(\matr{A}^{\dagger}\matr{A}\right)^{-1}_{(\mathit{N} \times \mathit{N})} \nonumber \\
    &= (\rm{SNR})_{\rm{full}}^{-2} \; \left(\displaystyle \frac{\mathit{N}}{2\mathit{p}\log{\mathit{N}}}\right) \; \frac{\mathit{t}_{\rm{int;full}}}{\mathit{t}_{\rm{cal}}} \; \frac{1}{\mathit{N}/2} \nonumber \\
    &= (\rm{SNR})_{\rm{full}}^{-2} \; \left(\displaystyle \frac{1}{\mathit{p}\log{\mathit{N}}}\right) \; \frac{\mathit{t}_{\rm{int;full}}}{\mathit{t}_{\rm{cal}}}
\end{align}
\noindent
When $t_{\rm{cal}}$ is large, the SNR of the reduced visibility matrix is larger and the corresponding gain variance is smaller. Assuming the case where the interval of calibration is same as the integration time in the FFT correlator and substituting the above equation into Equation~\ref{eq:visvar} we get the following for the variance in redundant visibilities:

\begin{equation}
    \label{eq:visvar_lowcadcal}
    (\rm{SNR})_{\rm{unique}}^{-2} \approx \displaystyle \frac{(\rm{SNR})_{\rm{full}}^{-2}}{\mathit{N}_{\alpha}} \left[1 + \left(\frac{2}{\mathit{p}\log{\mathit{N}}}\right)\right]
\end{equation}
\noindent
When redundant-baseline calibration is performed with the full visibility matrix, the gain covariance terms in Equation~\ref{eq:visvar} are around an order of magnitude smaller than the variance term so we drop the third term for clarity. In large arrays, where an FFT-correlator architecture would be preferable to an FX-correlator, the contribution of gain variance (second term) to the variance in calibrated redundant visibilities is smaller than the thermal noise in the measure visibilities (first term). Hence, the precision in gain variance obtained from using an $\mathcal{O}(N\log{N})$ calibrator is sufficient for the purpose of calibrating voltages for an FFT-correlator.

\subsubsection{Variance in Redundant Visibilities Calibrated using Subset Redundant Calibration}

In subset redundant calibration, the integration time does not change between the calibrator and the FFT-correlator. Consequently, the SNR of measured visibilities is the same for both data sets. However, the number of baselines used in the calibration process is lower, resulting in a higher variance in the estimated gains. For a calibrator that scales as $\mathcal{O}(N\log{N})$ with array size, the relationship between gain variance and number of baselines is given by Equation~\ref{eq:gainvar_subredcal} where the proportionality constant is the SNR of visibilities. Substituting Equation~\ref{eq:gainvar_subredcal} into Equation~\ref{eq:visvar} we get:

\begin{equation}
    \label{eq:visvar_subredcal}
    (\rm{SNR})_{\rm{unique}}^{-2} \approx \displaystyle \frac{(\rm{SNR})_{\rm{full}}^{-2}}{\mathit{N}_{\alpha}} \left[1 + \frac{2}{\mathit{p}\log{\mathit{N}}} + \frac{2}{\mathit{N}_{\alpha}} \sum\limits_{\mathit{(i,j)} \in \alpha} \rho_{\mathit{g}_\mathit{i}\mathit{g}_\mathit{j}}\right]
\end{equation}
\noindent
The gain covariance terms $\rho_{g_ig_j}$, for gains estimated using subset redundant calibration, are sometimes comparable to the gain variance terms. When the pre-factor (p) is small, the amplitude of the covariance scales similarly to the gain variance with increase in size of the array. This effectively doubles the contribution of the variance term in the above equation. When the pre-factor is large, the covariance is only a small fraction of the gain variance and can be ignored. Overall, the contribution of the second and third terms in the above equation is much smaller than the thermal noise in measured visibilities for any reasonably large array. Hence, the precision in gains estimated using subset redundant calibration is also sufficient for the purpose of calibrating voltages for an FFT-correlator. 

Figure~\ref{fig:scale_ants} might give a misleading impression that traditional redundant-baseline calibration is superior to either of the reduced redundant-baseline schemes, by providing gains that have orders-of-magnitude lower variance. However, this additional gain precision comes at a high computational cost and might not be necessary for large arrays where the contribution of gain variance to the overall scatter in redundant visibilities is only a small fraction of the thermal noise.

\subsection{Bias in Estimated Variables}

\begin{figure}
    \centering
    \includegraphics[width=\linewidth]{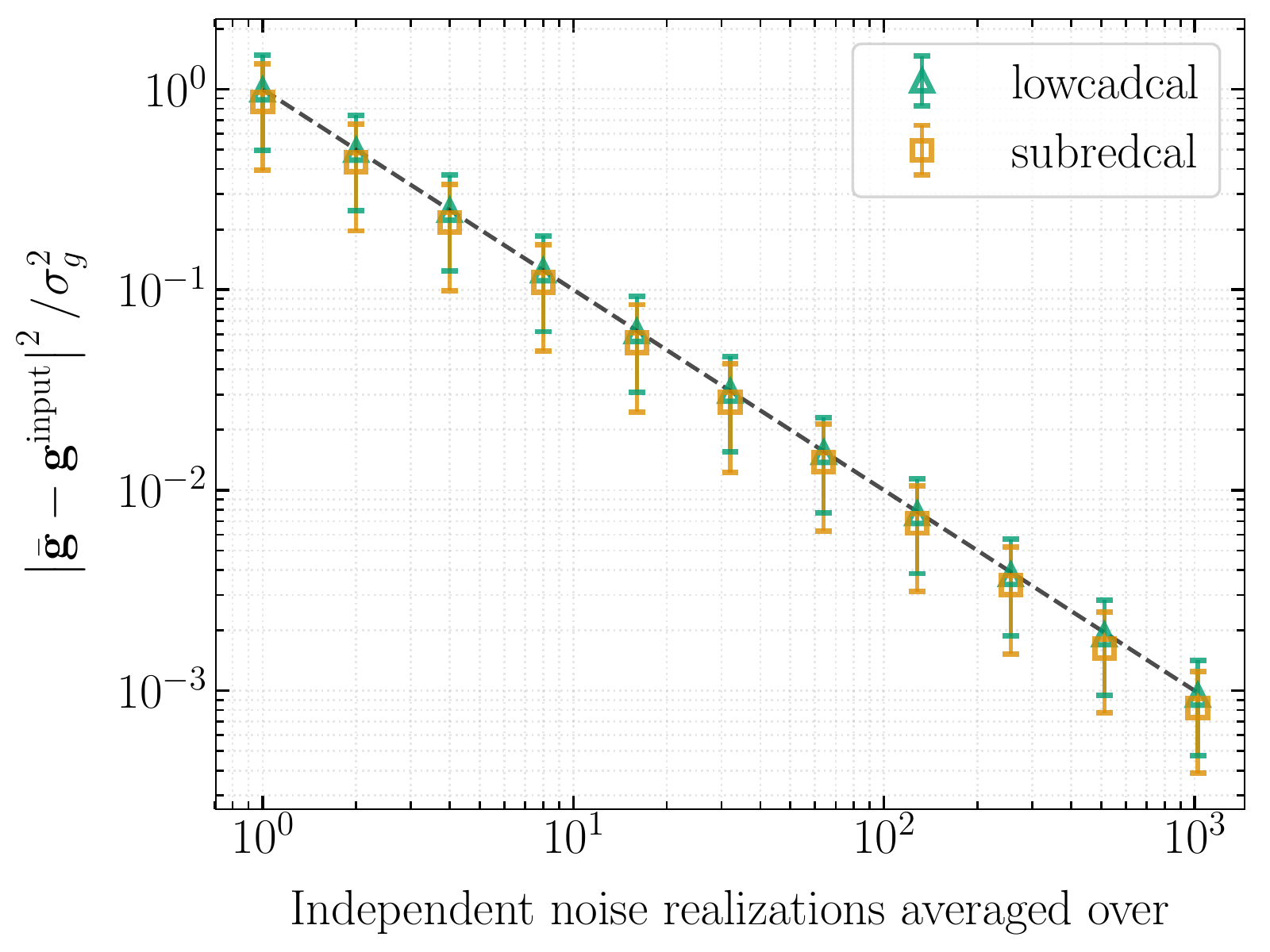}
    \caption{Deviation of averaged gains from their true value for low-cadence calibration (\texttt{lowcadcal}; green triangles) and subset redundant calibration (\texttt{subredcal}; orange squares). The markers are the result of a simulation with the same underlying input variables (both gains and unique visibilities) but different noise realisations of the measured visibilities. The x-axis shows the number of such simulations that the variables have been averaged over. The errorbars reflect the antenna-to-antenna variation. The dashed line represents the trend expected when the estimated gains differ from truth only within Gaussian noise. Both low-cadence calibration and subset redundant calibration yield unbiased gain solutions.
    \label{fig:gain_vis_bias}}
\end{figure}

The gains estimated using either reduced redundant-baseline calibration process are unbiased estimates of the true value. This can be verified through simulations that have constant underlying gains and visibilities, and different realisations of the noise in the measured visibilities. Averaging the solutions obtained over multiple such simulations decreases the noise in the estimated parameters and can expose an underlying bias, if any. Figure~\ref{fig:gain_vis_bias} shows the deviation in averaged gains from the input true gains, normalised by the variance expected in the gains. The errorbars represent the antenna-to-antenna variation which also averages down. Gains that have been averaged over $N_{\mathrm{sim}}$ independent noise realisations have a factor of $1/N_{\mathrm{sim}}$ smaller deviation, which is expected when the estimated gains differ from truth only within the Gaussian random noise in the measurement. This trend is marked by the dashed black line in the figure. A bias in gains, within the precision exposed by averaging down noise, would have resulted in a deviation from this trend.

The visibilities computed by the FFT-correlator ${\bf V^{unique}}$ are also unbiased when the visibility matrix used for reduced redundant-baseline has a high SNR. The deviation of the calibrated and redundant-baseline averaged visibilities from the simulated input, averages down according to the trend expected for Gaussian random noise. However, empirically, we find that the calibrated visibilities are sometimes biased when the variance in estimated gains is larger than $\sim$$10^{-5}$. It is possible that there is a low level of bias, that is not exposed by averaging over $N_{\rm{sim}}=4096$ simulations, in the visibilities calibrated with lower variance gains. To clarify, the gain variance plotted in all the figures in this paper has been normalised by the thermal noise in the visibilities used for calibration. The absolute gain variance in all these simulations is less than the empirical value above which we find a bias in the calibrated visibilities.

Assuming this threshold in gain variance is real, it translates to a requirement of an SNR $\gtrsim 20$ in the measured visibilities for a 300 antenna array when implementing low-cadence calibration, and higher when implementing subset redundant calibration. This minimum SNR requirement decreases as $1/\sqrt{N}$ with increase in array size and may not be an issue for large-N arrays.

This requirement of a minimum SNR in the reduced visibility matrix could be due two possible reasons-- (a) if the assumption that the non-linear equations of redundant-baseline calibration can be optimised by solving the linearized system of equations, does not hold at this limit or (b) if the product of gains form an asymmetric distribution about their mean value, and do not average down. If the former is true, a linearized solver of Equation~\ref{eq:redcal} would result in a biased solution when the noise in measured visibilities is high. However, we find that linearization based on Taylor expansion of variables and \texttt{omnical} always result in unbiased gains and visibilities irrespective of the SNR in the measured visibilities.

The more favourable explanation seems to be latter. As shown by \citet{odonoughue_and_moura_2012}, the probability distribution function (PDF) of the product of two complex Gaussian random variables is not a simple Gaussian distribution. Moreover, the resulting distribution can be asymmetric if two complex random variables are drawn from a non-zero mean Gaussian, as is the case with antenna gains. 

\subsection{$\chi_r^2$ of Estimated Gains and Visibilities}

\begin{figure}
    \centering
    \includegraphics[width=\linewidth]{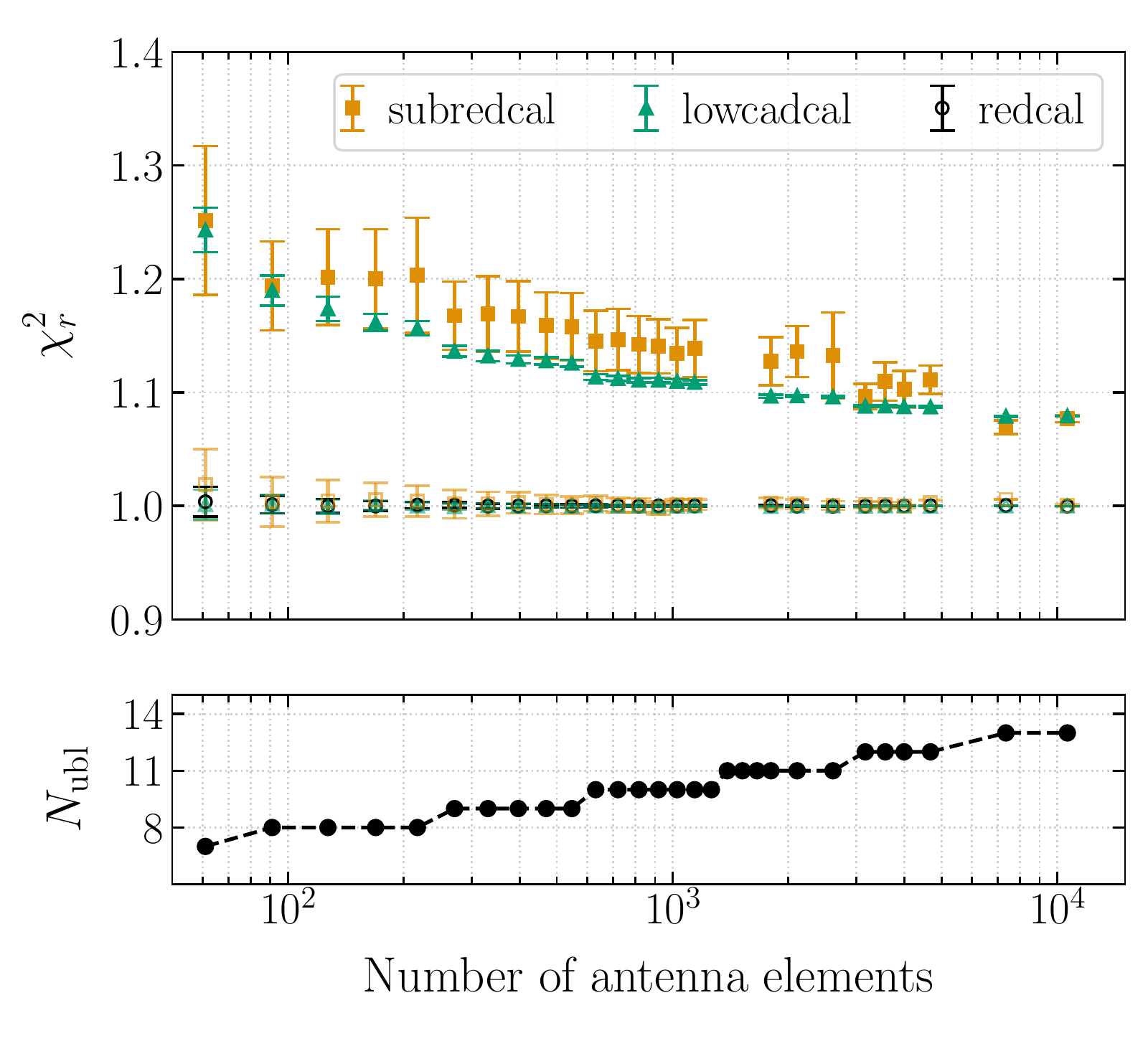}
    \caption{The upper panel shows the goodness-of-fit of the gains and unique visibilities that have been estimated using either reduced redundant-baseline calibration scheme. The hollow points (orange and green) show the fit of the estimated gains to the visibilities computed by the calibrator ${\bf V^{reduced}}$ and is close to one as expected. The hollow black points show the fit of gains estimated using full redundant-baseline calibration to ${\bf V^{full}}$. The solid orange and green points show the $\chi_r^2$ of the fit to the full visibility matrix ${\bf V^{full}}$ and these are also not far from one. The associated errorbars show the range of $\chi_r^2$ obtained over multiple simulations. The jump in $\chi_r^2$ for low-cadence calibration arises from a non-uniform increase in the calibrator size. The number of unique baseline-types that are cross-correlated by the calibrator are shown in the lower panel as a proxy for the size of the calibrator.
    \label{fig:chisq_scale_ants}}
\end{figure}

As was discussed in Section~\ref{sec:redredcal:scatter}, the gains computed using a reduced redundant-baseline calibration scheme, are estimated from the visibility matrix ${\bf V^{reduced}}$ but applied to a different visibility matrix ${\bf V^{unique}}$. The redundant-baseline calibration process is designed to minimise the $\chi_r^2$ between the visibility matrix used for calibration and the estimated variables. In the case of reduced redundant-baseline calibration, this is the $\chi_r^2$ evaluated between the estimated gains, model visibilities that are discarded by the calibrator and the reduced visibility matrix. In Figure~\ref{fig:chisq_scale_ants}, this $\chi_r^2$ is represented by hollow orange and green markers for subset redundant calibration and low-cadence calibration respectively, and has the expected value of one.

The $\chi_r^2$ estimated using gains computed by the reduced redundant-baseline calibration process, unique visibilities computed by the FFT-correlator and the full visibility matrix is a better metric to assess the effectiveness of the calibration process. This $\chi_r^2$ is represented by the solid orange and green points in Figure~\ref{fig:chisq_scale_ants}, for subset redundant calibration and low-cadence calibration respectively. The $\chi_r^2$ for array sizes with number of antennas in the range 1500--5000 is estimated over only 16 simulations, rather than 256, due to long simulation times. For the same reason, the $\chi_r^2$ for the two largest array sizes with $N>5000$ antennas is estimated using only 2 simulations.

Both the reduced redundant-baseline calibration schemes yield a $\chi^2_r$ that is close to one, indicating that the estimated parameters are a reasonable fit to the full visibility matrix. The larger $\chi^2_r$ of subset redundant calibration as compared to low-cadence calibration, and the larger range of $\chi_r^2$ obtained over multiple simulations (represented by the errorbars) is a direct consequence of the higher gain variance in the former compared to the latter. The non-smooth trend in the $\chi_r^2$ of low-cadence calibration is due to a non-uniform increase in the size of the calibrator used in simulation. The jump in $\chi_r^2$ of low-cadence calibration is correlated with the increase in number of unique-baseline types processed by the calibrator, because this leads to a jump in the integration time available to each cycle of correlation in the calibrator. 

\begin{figure}
    \centering
    \includegraphics[width=\linewidth]{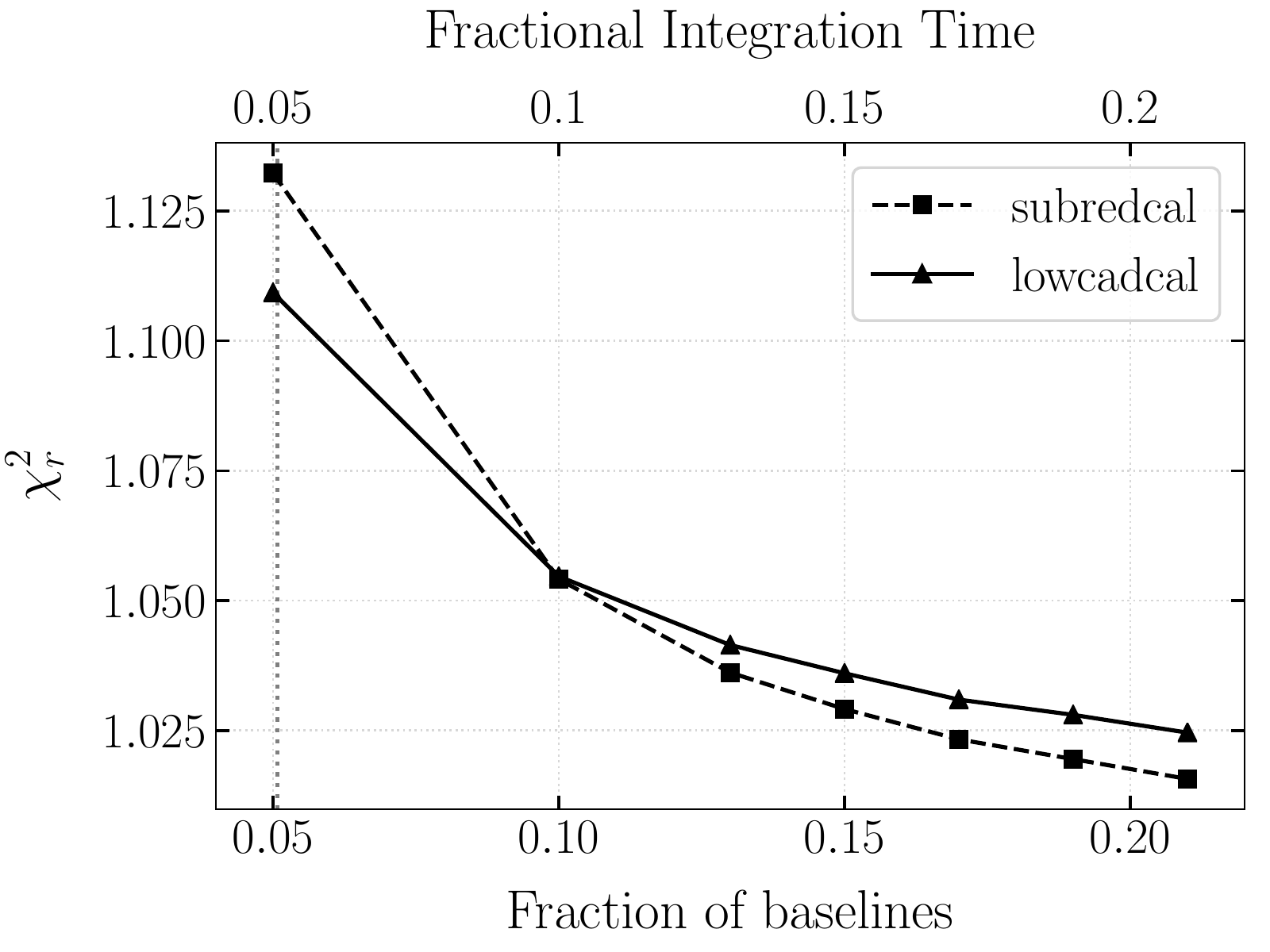}
    \caption{Reduced chi-squared as a function of the computational resources allocated to a reduced redundant calibrator for a 331\,antenna array. The size of the calibrator is marked in the fraction of baselines that can be processed for subset redundant calibration (bottom axis) or the fraction of integration time that can be spent on each baseline for low-cadence calibration (top axis). The vertical dotted line marks the size of a $\mathcal{O}(N \log{N})$ calibrator. At larger calibrator sizes, subset redundant calibration yields better gain estimates than low-cadence calibration.
    \label{fig:low_v_sub}}
\end{figure}

Figure~\ref{fig:low_v_sub} compares the performance of low-cadence calibration to subset redundant calibration for calibrators of various sizes operating a fixed array size. When the resources allocated to the calibrator are sufficient to cross-correlate exactly $N\log{N}$ baselines, which was the assumption throughout this paper, low-cadence calibration yields better gain estimates. This calibrator size is marked by a vertical dotted line in the figure. However, at larger calibrator sizes, subset redundant calibration performs better than low-cadence calibration. This is because the subset redundant calibrator preferentially spends time on baselines that have a higher constraining power and integrates them to a higher SNR than a low-cadence calibrator of a similar size. 

\section{Conclusion}
\label{sec:conclusion}

Low frequency radio interferometers with $N\gg 1000$ antennas are being proposed for targeting the 21\,cm signal at cosmological distances. These arrays will face a significant computational cost in building traditional FX-correlators that scale as $\mathcal{O}(N^2)$ with the number of antennas in the array. FFT-correlators attempt to decrease this scaling to $\mathcal{O}(N\log{N})$, thereby decreasing the cost of the correlator backend. However, FFT-correlators only produce meaningful output when the input antenna voltages are calibrated. In the past, the few experiments that have used FFT-correlators on redundant arrays had an FX-correlator working in parallel for computing the visibilities required for calibration. This is, however, a non-scalable solution for large-N arrays. In this paper we propose a $\mathcal{O}(N\log{N})$ calibrator design that can operate in parallel with the FFT-correlator, forming a self-contained correlator system that can be scaled to large-N arrays built on a regular grid. We discuss two calibration algorithms, that can be employed by such a calibrator, which are modifications of redundant-baseline calibration. 

Redundant-baseline calibration relies on the array having multiple redundant baselines or antenna pairs with the same displacement vector. Non-redundancies in the array, originating from beam variation, antenna placement errors, etc. could lead to errors in the gain solution as discussed by~\citet{orosz_et_al2018} and~\citet{Byrne_et_al_2019}. These gain errors are likely to affect low-cadence calibration and subset redundant calibration, even though the FFT-correlator averages theoretically-redundant baselines together. An analysis of the impact of non-redundancy on an FFT-correlated telescope is deferred for future work.

In low-cadence calibration, antenna pairs are cross-correlated in a round-robin fashion over multiple integration cycles, producing the full visibility matrix once every given number of cycles. However, the time period between two calibration cycles $t_{\rm{cal}}$ is fixed by the inherent gain variability of the array. This leads to a decrease in the integration time available for each cycle as the array size increases. A decrease in integration time results in lower SNR visibilities within the calibrator, and consequently a higher variance in the estimated gains. For a calibrator that is restricted to an $\mathcal{O}(N\log{N})$ scaling, the gain variance scales as $1/\log{N}$ with increasing array size. The contribution of this gain variance to the overall variance in the visibilities computed by the FFT-correlator is much smaller than the contribution of thermal noise in measurements. Hence, low-cadence calibration is a suitable calibration scheme to estimate gains that are effective in calibrating voltages for the FFT-correlator.

In subset redundant calibration, the calibrator computes the visibilities of only a few baseline groups without compromising on their SNR. By nature of the system of equations solved by redundant-baseline calibration, the larger baseline groups contribute higher constraining power to antenna gains. Subset redundant calibration exploits this property and computes cross-correlations of only these baseline groups. The minimum number of baseline-types that need to be considered for subset redundant calibration is determined by the null space of the solution set, which should ideally not have more than the four known degeneracies. However, using only a small fraction of all the baselines, that satisfy this degeneracy criterion, could still result in antenna gains that have a high variance and covariance. A calibrator that can cross-correlate $N\log{N}$ baselines, results in antenna gains that have a $1/\log{N}$ scaling in variance as well. As in the case of low-cadence calibration, the variance of the estimated antenna gains forms only a small fraction of the total scatter in visibilities computed by the FFT-correlator. Hence, subset redundant calibration is also a suitable scheme of calibration for estimating gains that can minimise scatter in redundant visibilities.

The gains, estimated using either reduced redundant-baseline calibration method, are unbiased and converge to the true value when noise averaged. When comparing low-cadence calibration and subset redundant calibration, we find that low-cadence calibration consistently yields lower variance antenna gains when the size of the calibrator is held constant. This consequently leads to a better fit to measured visibilities and the $\chi^2_r$ estimated using the variables computed by low-cadence calibration is lower than that of subset redundant calibration. However, for large-N arrays low-cadence calibration could involve optimising over a few hundred thousand equations which could potentially hamper the real-time nature of the calibration parameters required. If the calibrator for a given array has more computational resources, subset redundant calibration can result in better antenna gain estimates. Ultimately, the calibration method that is suitable for a large array depends on the antenna design parameters, the array layout, computational resources available and the science goal at hand.


\section*{Acknowledgements}
We would like to thank Jack Hickish, Phil Bull, Adrian Liu, Nithyanandan Thyagarajan, Adam Beardsley, and Judd Bowman for valuable discussions that contributed to this paper. We would also like to thank the anonymous reviewer for providing suggestions that improved the readability and value of this paper. This material is based upon work supported by the National Science Foundation under grants \#1636646 and \#1836019. This research is funded in part by the Gordon and Betty Moore Foundation. JSD gratefully acknowledges the support of the NSF AAPF award \#1701536 and the Berkeley Center for Cosmological Physics.

\section*{Data Availability}
The data underlying this article has been generated using code publicly available in the repository: \url{https://github.com/HERA-Team/hera_sim}.


\bibliographystyle{mnras}
\bibliography{refs_redredcal}



\appendix
\section{Hierarchical Calibration}
\label{app:hierarcal}
Hierarchical redundant-baseline calibration was originally presented by \citet[appendix~B]{zheng_et_al2014}. It is based on separating a redundant array into sub-arrays and cross-correlating all the antenna pairs in each sub-array. Antenna gains are estimated by performing redundant-baseline calibration on each sub-array independently. The degenerate parameters of each sub-array are tied together by choosing one antenna from each sub-array and redundantly calibrating the array they form. For redundant arrays where the correlator also operates hierarchically \citep{tegmark_and_zaldarriaga2010}, such a calibrator system would parallel the correlator layout in the field and reduce networking. However, this is not an optimal calibration solution for FFT-correlators that operate on the entire array.

The spatial Fourier transform in an FFT-correlator can only be performed on a machine containing the voltages of all antennas in the array. This requires a large corner-turn as in FX-correlators, where the voltages from all antennas over a narrow bandwidth are collected at a central server using an Ethernet switch or a similar device. When data from all the antennas is available at a central location, the baselines that yield optimal gain solutions can be chosen in any fashion for the purpose of calibration. While hierarchical calibration has the potential to simplify the networking required for large arrays, the networking required for FFT-correlators renders this simplification moot. For this reason, hierarchical calibration has been classified as a special case of subset redundant calibration and is not presented in more detail.


\bsp	
\label{lastpage}
\end{document}